\documentclass[preprint,aps,12pt,showpacs,nofootinbib,tightenlines,amsmath,amssymb]{revtex4}
\usepackage{amsmath}
\usepackage{graphicx}
\usepackage{amssymb}
\usepackage{color}

\newcommand{\btd}{\bigtriangledown}
\newcommand{\be}{\begin{eqnarray}}
\newcommand{\ee}{\end{eqnarray}}

\newcommand{\Tr}{\mathop{\rm Tr}\nolimits}

\newcommand{\ob}{\bar{\omega}}
\newcommand{\ot}{\tilde{\omega}}

\newcommand{\Gmt}{\tilde{\mathsf \Gamma}}
\newcommand{\Gm}{\mathsf \Gamma}
\newcommand{\Gmct}{\tilde{\varGamma}}
\newcommand{\Gmc}{\varGamma}

\newcommand{\Lm}{\mathsf L}

\newcommand{\Sm}{\mathsf S}
\newcommand{\Rm}{\mathsf R}
\newcommand{\Rmt}{\tilde{\mathsf R}}

\newcommand{\Rct}{\tilde{\cal R}}
\newcommand{\wb}{\bar{w}}
\newcommand{\pb}{\bar{\varphi}}
\newcommand{\ab}{\bar{a}}

\newcommand{\gb}{\bar{g}}
\newcommand{\chib}{\bar{\chi}}
\newcommand{\Ob}{\bar{\Omega}}
\newcommand{\Om}{\mathsf \Omega }
\newcommand{\Omt}{\tilde{{\mathsf \Omega }} }
\newcommand{\td}{\bigtriangledown}

\newcommand{\chih}{\hat{\chi}}

\newcommand{\Nh}{\hat{N}}
\newcommand{\mk}{m_{\kappa}}
\newcommand{\lk}{l_{\kappa}}
\newcommand{\ek}{\eta_{\kappa}}
\newcommand{\Mk}{H_{\kappa}}
\newcommand{\Lk}{L_{\kappa}}
\newcommand{\kh}{\hat{\kappa}}

\begin{document}
\def\intdk{\int\frac{d^4k}{(2\pi)^4}}
\def\sla{\hspace{-0.17cm}\slash}
\hfill


\title{Theory of Quantum Gravity Beyond Einstein and \\ Space-time Dynamics with Quantum Inflation\footnote{This article is based on the invited talk presented at  the International Conference on Gravitation and Cosmology/the fourth Galileo-Xu Guangqi Meeting and the recent progresses made in ref.\cite{YLW} .}}

\author{Yue-Liang Wu} 
\affiliation{State Key Laboratory of Theoretical Physics(SKLTP), Kavli Institute for Theoretical Physics China (KITPC) \\
Institute of Theoretical Physics, Chinese Academy of Sciences, Beijing, 100190, China\\
International Centre for Theoretical Physics Asia-Pacific (ICTP-AP) \\
University of Chinese Academy of Sciences (UCAS) }


\begin{abstract}

In this talk, I present a theory of quantum gravity beyond Einstein. The theory is established based on spinnic and scaling gauge symmetries by treating the gravitational force on the same footing as the electroweak and strong forces. A bi-frame space-time is initiated to describe the laws of nature. One frame space-time is a globally flat coordinate Minkowski space-time that acts as an inertial reference frame for the motions of fields, the other is a locally flat non-coordinate Gravifield space-time that functions as an interaction representation frame for the degrees of freedom of fields. The Gravifield is sided on both the globally flat coordinate space-time and locally flat non-coordinate space-time and characterizes the gravitational force. Instead of the principle of general coordinate invariance in Einstein theory of general relativity, some underlying principles with the postulates of coordinate independence and gauge invariance are motivated to establish the theory of quantum gravity. When transmuting the Gravifield basis into the coordinate basis in Minkowski space-time, it enables us to obtain equations of motion for all quantum fields and derive basic conservation laws for all symmetries. The gravity equation is found to be governed by the total energy-momentum tensor defined in the flat Minkowski space-time. When the spinnic and scaling gauge symmetries are broken down to a background structure that possesses the global Lorentz and scaling symmetries, we arrive at a Lorentz invariant and conformally flat background Gravifield space-time that is characterized by a cosmic vector with a non-zero cosmological mass scale. We also obtain the massless graviton and massive spinnon. The resulting universe is in general not isotropic in terms of conformal proper time and turns out to be inflationary in light of cosmic proper time. The conformal size of the universe has a singular at the cosmological horizon to which the cosmic proper time must be infinitely large. We show a mechanism for quantum inflation caused by the quantum loop contributions. The Gravifield behaves as a Goldstone-like field that transmutes the local spinnic gauge symmetry into the global Lorentz symmetry, which makes the spinnic gauge field becomes a hidden gauge field. As a consequence, the bosonic gravitational interactions can be described by the Goldstone-like Gravimetric field and space-time gauge field. The Einstein theory of general relativity is expected to be an effective low energy theory.  Two types of gravity equation are resulted, one is the extension to Einstein's equation of general relativity, and the other is a new type of gravitational equation that characterizes the spinnon dynamics.  

\end{abstract}
\pacs{04.60.-m, 11.15.-q, 04.90.+e}

\maketitle

\section{introduction}

One hundred years ago, when Einstein proposed the theory of general relativity as a natural extension of special relativity, its foundation was based on the postulate: {\it the laws of physics must be of such a nature that they apply to system of reference in any kind of motion}.  Namely, {\it the physical laws of nature are to be expressed by equations which hold good for all systems of coordinates}\cite{GR}.  Such a postulate leads the gravitational force to be characterized by a Riemann geometry in a curved space-time. Namely the gravitational force lies in the dynamic of metric field $g_{\mu\nu}(x)$ in Riemannian geometry of curved space-time. Thus the physical laws, in the theory of general relativity, are invariant under the general linear transformations of local GL(4,R) symmetry, which indicates that space and time cannot be well defined in such a way that the differences of the spacial coordinates or time coordinates can be directly measured by the standard ways proposed in special relativity. 

On the other hand, the relativistic quantum field theory(QFT) has successfully unified the quantum mechanics and special relativity and provided a remarkable theoretical framework to describe the microscopic world.  The three basic forces,  i.e., electromagnetic, week and strong interactions, have successfully been described by the so-called standard mode (SM) of gauge interactions, which was established based on the gauge symmetries\cite{YM} of quarks and leptons within the framework of QFT\cite{QED1, QED2, QED3, QED4,EW1,EW2,EW3,QCD1,QCD2,QCD3}  in the flat Minkowski space-time. The SM has been tested by more and more precise experiments, the advent of the SM-like Higgs boson at the LHC \cite{higgs2012a,higgs2012c} motivates us to study more fundamental theory including quantum gravity. Einstein theory of general relativity\cite{GR} as the theory of gravity has well been applied to describe the macroscopic world.  The gravitational force is characterized as the dynamics of a metric in the curved Riemannian space-time, which is in contrast to three basic forces of gauge interactions. Such an odd dichotomy becomes the obstacle for a unified description of gravity with three basic forces governed by gauge symmetries, and causes the difficulty for the quantization of gravitational force. It is believed that a QFT description for the gravitational force is crucial to understand the origin of universe, such as the singularity and inflation of early universe\cite{IU1,IU2,IU3}(for a recent review, see,e.g. Ref.{\cite{IU4}, and reference there in).

To make a significant simplification for the equations of motion, it was emphasized by Einstein in his original paper that: ``if  $-\det g_{\mu\nu} $ (determinant of metric in curved coordinate space-time) is always finite and positive, it is natural to settle the choice of coordinates a {\it posteriori } in such a way that this quantity is always equal to unity." ``Thus, with this choice of coordinates, only substitutions for which the determinant is unity are permissible. " ``But it would be erroneous to believe that this step indicates a partial abandonment of the general postulate of relativity. We do not ask : what are the laws of nature which are covariant in face of all substitutions for which the determinant is unity? but our question is: what are the general covariant laws of nature?  " Alternatively, we shall make a sensible principle for the gravitational interaction within the framework of QFT and do ask what are the laws of nature which are covariant in face of all substitutions for which the determinant is unity.

The symmetry principle is the well established principle, which states that the basic forces of elementary particles are governed by gauge symmetries and characterized by quantum fields defined in the flat Minkowski space-time of coordinates.  Thus we shall treat the gravitational force on the same footing as the electroweak and strong forces and introduce relevant gauge symmetries to characterize the gravitational interaction. As all symmetry groups in the SM are related to the intrinsic quantum numbers of quarks and leptons. The relevant gauge symmetry for the gravitational force is thought to correlate with the spinnic Lorentz symmetry group SO(1,3)$\cong$ SP(1,3) of quarks and leptons, which should distinguish from the global Lorentz symmetry group SO(1,3) of coordinates in the Minkowski space-time. The notations SP(1,3) and SO(1,3) are used to make such a distinction.  Some early papers and review articles on gauge theories of gravity may be found in \cite{GGT1,GGT2,GGT3,GGT4,GGT5,GGTR1,GGTR2,GGTR3}.

In the present consideration, the QFT description of all basic forces including gravity must obey, in addition to the well established principles for the special relativity and quantum mechanics, some underlying principles alternative to the Einstein's general coordinate invariance. Such a theory of quantum gravity is based on the postulates/principles:  
 (i) a bi-frame space-time is needed to describe the laws of nature. One frame space-time is a globally flat coordinate Minkowski space-time that acts as an inertial reference frame for the motions of fields, the other is a locally flat non-coordinate Gravifield space-time that functions as an interaction representation frame for the degrees of freedom of fields;  (ii) the kinematics of all fields obeys the principles of special relativity and quantum mechanics; (iii) the dynamics of all fields is characterized by basic interactions governed by gauge symmetries; (iv)  the theory must be invariant under the local spinnic and scaling gauge transformations for fields as well as under the global Lorentz and scaling transformations for coordinates in Minkowski space-time; (v) a gauge invariant theory must be expressed in a coordinate-independent form in Gravifield space-time.

\section{Spinnic and Scaling Gauge Symmetries for Gravity Theory}

Quarks and leptons as the basic building blocks of nature in the SM carry out the intrinsic quantum numbers of electric charges, isospin charges and color charges. These intrinsic quantum numbers of charges are correlated through gauge symmetries U(1)$_Y$ x SU(2)$_L$ x SU(3)$_c$. It is those symmetries that govern three basic forces, electromagnetic, week and strong interactions, respectively.  All quarks and leptons as fermionic quantum fields $\Psi(x)$ are known to carry out spinnic quantum numbers.  Within the framework of QFT, the Lorentz invariance of theory under the global Lorentz transformation of coordinates with the symmetry group SO(1,3) requires the spinnic quantum numbers of quarks and leptons to be correlated through the spinnic Lorentz symmetry group SP(1,3) $\cong$ SO(1,3).  

In establishing a gauge theory of gravitational interaction, motivated from the gauge symmetries U(1)$_Y$ x SU(2)$_L$ x SU(3)$_c$ for the basic forces of electromagnetic, weak and strong interactions, it is natural to gauge the spinnic symmetry SP(1,3) of quarks and leptons. So that the gauge symmetry SP(1,3) governs a basic force of {\it spinnic gauge interaction}. In analogous to the introduction of gauge fields ${\cal A}_{\mu}(x)$ for gauge symmetries U(1)$_Y$ x SU(2)$_L$ x SU(3)$_c$,  the corresponding spinnic gauge field and field strength for the gauge symmetry SP(1,3) are introduced 
\begin{eqnarray}
& & \Om_{\mu}(x) = g_s\; \Om_{\mu}^{ab}(x)\,  \frac{1}{2} \Sigma_{ab} \; , \quad  \Sigma^{ab} = \frac{i}{4}[\gamma^a, \gamma^b]  \nonumber \\
& & {\cal R}_{\mu\nu}(x) =  \partial_{\mu} \Om_{\nu}(x) - \partial_{\nu} \Om_{\mu}(x) - i [\Om_{\mu}(x) ,  \Om_{\nu}(x) ] = 
g_s {\cal R}_{\mu\nu}^{ab} (x)\,  \frac{1}{2} \Sigma_{ab}\; , 
\end{eqnarray}
with $g_s$ the coupling constant and $\Sigma^{ab}$ the generators of spinnic group SP(1,3) in the spinor representation,
\begin{eqnarray}
& & [\Sigma^{ab}, \Sigma^{cd}] = i (\Sigma^{ad}\eta^{bc} -\Sigma^{bd}  \eta^{ac} - \Sigma^{ac} \eta^{bd} + \Sigma^{bc} \eta^{ad}) , \nonumber \\
& & [ \Sigma^{ab}, \gamma^c] = i ( \gamma^a \eta^{bc} - \gamma^b\eta^{ac} ).
\end{eqnarray}

To ensure the theory be invariant under both the spinnic gauge transformation for quantum fields and global Lorentz transformation for coordinates, it is a necessity to introduce a bi-covariant vector field $\hat{\chi}_a^{\;\;\mu}(x)$ sided on a locally flat non-coordinate space-time and valued in a vector representation of Lorentz group SO(1,3) in the flat Minkowski space-time, so that we are able to define the Lorentz invariant derivative
\be
\hat{\chi}_a = \hat{\chi}_a^{\;\;\mu}(x)\partial_{\mu} = \eta_{\mu\nu} \hat{\chi}^{\;\;\mu}_a(x) \partial^{\nu}\; ,
\ee

For illustration, let us construct a gauge invariant action for Dirac fermions which can be quarks and leptons.  The action is supposed to be invariant under the local spinnic gauge transformations for the quantum fields, and also under the global Lorentz and scaling transformations for the coordinates
\begin{eqnarray}
\label{actionF}
S_{F}  & = & \int d^{4}x\; \chi(x)\, \frac{1}{2} \{ \bar{\Psi}(x) \hat{\chi}^{\mu}(x) i{\mathcal D}_{\mu} \Psi(x) + H.c.\}  
\end{eqnarray}
where the covariant derivative for the usual internal gauge and spinnic gauge symmetries is given by
\begin{eqnarray}
 i {\mathcal D}_{\mu} & = & i\partial_{\mu}  + {\cal A}_{\mu}(x) + \Om_{\mu}(x) = i D_{\mu} + \Om_{\mu}(x)  \; ,\quad  i D_{\mu}  = i\partial_{\mu}  + {\cal A}_{\mu}(x) \, , 
 \end{eqnarray}
 and the bi-covariant vector field $\hat{\chi}_a^{\;\;\mu}(x)$ is a necessary ingredient 
 \begin{eqnarray}
 \hat{\chi}^{\mu}(x)i {\mathcal D}_{\mu} & \equiv &  \frac{1}{2} \gamma^a \hat{\chi}_a^{\;\;\mu}(x)  i {\mathcal D}_{\mu} = \frac{1}{2} \gamma_a \eta^{ab} \hat{\chi}_{b\nu}(x)  \eta^{\mu\nu}   i {\mathcal D}_{\mu}  \; , 
\end{eqnarray}
so that the above action becomes invariant under the usual internal and spinnic gauge transformations as well as under the global Lorentz and scaling transformations in the flat Minkowski space-time. It is easy to show that the Hermiticity of the action automatically ensures the invariance under the scaling gauge transformation.

The quantum fields $\Psi_n(x)$ $(n=1,2,\cdots)$ denote Dirac fermions, such as quarks and leptons in SM. ${\cal A}_{\mu}(x) = g_A {\cal A}_{\mu}^I(x) T^I$ ($I=1,2,\cdots$) with $g_A$ the coupling constant represent gauge fields with $T^I$ the generators of symmetry group $G$ which can be G=U(1)$_Y$ x SU(2)$_L$ x SU(3)$_c$ in SM. The generators satisfy the group algebra $[T^I, T^J] = i f^{IJK} T^K$  with the trace normalization $tr T^I T^J = \frac{1}{2} \delta^{IJ}$. The Greek alphabet ($\mu,\nu = 0,1,2,3$) and  Latin alphabet ($a,b,=0,1,2,3$)  are adopted to distinguish four-vector indices in coordinate space-time and non-coordinate space-time, respectively.  Both the Geek and Latin indices are raised and lowered by the constant metric matrices, i.e., $\eta^{\mu\nu} $ and $\eta^{ab}$ with the signature $-2$, $\eta^{\mu\nu} $ or $\eta_{\mu\nu} = diag.(1,-1,-1,-1)$ and $\eta^{ab}$ or $\eta_{ab} = diag. (1,-1,-1,-1)$. The system of units is chosen such that $c = \hbar = 1$.

Explicitly, the action can be shown to be invariant under the internal gauge transformation $g(x) \in G$,
\begin{eqnarray}
& & {\cal A}_{\mu}(x) \to {\cal A}'_{\mu}(x) = g(x) {\cal A}_{\mu}(x) g^{\dagger}(x) +   g(x) i\partial_{\mu} g^{\dagger}(x) \; , \nonumber \\
& & \Psi(x) \to \Psi'(x) = g(x) \Psi(x)
\end{eqnarray}
and the spinnic gauge transformation $S(x) = e^{i\alpha_{ab}(x) \Sigma^{ab}/2}\in $ SP(1,3),
\begin{eqnarray}
& & \Om_{\mu}(x) \to \Om'_{\mu}(x) = S(x) \Om_{\mu}(x) S^{-1}(x) + S(x)i \partial_{\mu} S^{-1}(x), \nonumber \\
& &\Psi(x) \to \Psi'(x) = S(x) \Psi(x) \; , \quad S(x) \gamma^{a} S^{-1}(x) = \Lambda^a_{\;\;\; b}(x)\; \gamma^b\, .
\end{eqnarray}
as well as under a local scaling gauge transformation 
\begin{eqnarray}
& & \Psi(x) \to \Psi'(x) = \xi^{3/2} (x) \Psi(x) \; .
\end{eqnarray}
While the internal gauge fields $ {\cal A}_{\mu} (x)$ and spinnic gauge field  $\Om_{\mu}(x)$ are unchanged in the local scaling gauge transformation. The bi-covariant vector field $\hat{\chi}_a^{\;\;\mu}(x)$ transforms as 
\begin{eqnarray}
& & \hat{\chi}_a^{\;\; \mu}(x) \to  \hat{\chi}_a^{'\;\mu}(x) =  \Lambda_a^{\;\; b}(x) \hat{\chi}_b^{\;\; \mu}(x) \; , \nonumber \\
& & \hat{\chi}_a^{\;\; \mu}(x) \to  \hat{\chi}_a^{'\; \mu}(x) = \xi(x) \hat{\chi}_a^{\;\; \mu}(x) \, ,
\end{eqnarray}
under the local spinnic  and scaling gauge transformations, respectively.

Under the global Lorentz and scaling transformations, the gauge fields, bi-covariant vector field and Dirac fermions transform in the vector and spinor representations of Lorentz group in the flat Minkowski space-time,
\begin{eqnarray}
& & x^{\mu} \to x^{'\mu} = L^{\mu}_{\; \;\; \nu}\; x^{\nu}\; , \, A_{\mu} (x) \to A'_{\mu}(x') = L_{\mu}^{\; \;\; \nu}\; A_{\nu}(x)  \; , \,  \hat{\chi}_a^{\;\; \mu}(x) \to  \hat{\chi}_a^{'\;\mu}(x') = L^{\mu}_{\; \;\; \nu}\; \hat{\chi}_a^{\;\; \nu}(x)\, , \nonumber \\
& & x^{\mu} \to x^{'\mu} = \lambda^{-1}\; x^{\mu}, \quad {\cal A}_{\mu} (x) \to {\cal A}'_{\mu}(x') = \lambda\; {\cal A}_{\mu}(x), \quad
\Psi(x) \to \Psi'(x') = \lambda^{3/2}\; \Psi(x), \nonumber \\
& & \Om_{\mu}(x) \to \Om'_{\mu}(x')  = \lambda\; \Om_{\mu}(x) \; , \,    \hat{\chi}_a^{\;\; \mu}(x) \to  \hat{\chi}_a^{'\; \mu}(x') = \hat{\chi}_a^{\;\;\mu}(x)\, .
\ee
The transformations satisfy the orthogonal conditions
\be
& & \Lambda_a^{\;\; b}(x) \Lambda_c^{\;\; d}(x) \eta^{ac} = \eta^{bd}\; , \quad \Lambda_a^{\;\; b}(x) \in SP(1,3)\; , \nonumber \\
& & L^{\mu}_{\; \;\; \nu} L^{\rho}_{\; \;\; \sigma} \eta_{\mu\rho} = \eta_{\nu\sigma}\; , \qquad \quad  L^{\mu}_{\; \;\; \nu} \in SO(1,3)\; .
\ee

Let us introduce a bi-covariant vector field $\chi_{\mu}^{\;\; a} (x)$ which is dual to  $ \hat{\chi}^{\;\;\mu}_a(x)$ with the following orthonormal conditions  
\begin{eqnarray}
\chi_{\mu}^{\;\; a} \hat{\chi}^{\;\;\nu}_a(x) = \chi_{\mu\, a} (x) \hat{\chi}_b^{\;\; \nu}(x)  \eta^{ab} = \eta_{\mu}^{\;\;\nu} \; ,\quad 
\hat{\chi}^{\;\; \mu}_b(x) \chi_{\mu}^{\;\; a} (x)  =   \hat{\chi}_{a\, \mu}(x)\chi_{\nu}^{\;\; b} (x) \eta^{\mu\nu} = \eta_{a}^{\;\;b} \; .
\end{eqnarray}
which holds as long as  the determinant of $\hat{\chi}_{a}^{\;\; \mu}(x)$ is nonzero, namely
\begin{eqnarray}
\det \chi_{\mu}^{\;\; a}(x)\equiv \chi(x) = \frac{1}{\det \hat{\chi}_a^{\;\;\mu}(x) }\equiv \frac{1}{\hat{\chi}(x)},\quad \hat{\chi}(x)\equiv \det \hat{\chi}_{a}^{\;\; \mu}(x)\neq 0  \, . 
\end{eqnarray}
The bi-covariant vector field $\chi_{\mu}^{\;\; a}(x)$ transforms as 
\be
& &  \chi_{\mu}^{\;\; a}(x) \to \chi_{\mu}^{'\; a}(x) = \Lambda^{a}_{\;\; b} (x) \chi_{\mu}^{\;\; b}(x) \; , \nonumber  \\
& &  \chi_{\mu}^{\;\; a}(x) \to  \chi_{\mu}^{'\; a}(x) = \xi^{-1}(x) \chi_{\mu}^{\;\; a}(x)
\ee
under the local spinnic and scaling gauge transformations, respectively, and 
\begin{eqnarray}
& &  \chi_{\mu}^{\;\; a} (x) \to  \chi_{\mu}^{'\; a}(x') = L_{\mu}^{\; \; \nu}\, \chi_{\nu}^{\;\; a} (x')\, , \quad 
 x^{\mu} \to x^{'\mu} = L^{\mu}_{\; \;\; \nu}\; x^{\nu}\; , \nonumber \\
& & \chi_{\mu}^{\;\; a}(x) \to  \chi_{\mu}^{'\;a} (x') = \chi_{\mu}^{\;\; a} (x')\, , \quad 
 x^{\mu} \to x^{'\mu} = \lambda^{-1}\; x^{\mu}\; ,
\end{eqnarray}
under the global Lorentz and scaling transformations, respectively.  

$ \chi_{\mu}^{\;\; a} (x) $ can be regarded as a bi-covariant vector field sided on the globally flat Minkowski space-time and valued in the vector representation of SP(1,3) in the locally flat non-coordinate space-time,
\be
\chi_{\mu} =  \chi_{\mu}^{\;\; a} (x)\, \frac{1}{2} \gamma_a \, .
\ee
We would like to address that the bi-covariant vector field $\chi_{\mu}^{\;\; a} (x)$ or dual bi-covariant vector field $ \hat{\chi}_a^{\;\;\mu}(x)$ transforms in a covariant form on both the globally flat Minkowski space-time of coordinates and the locally flat non-coordinate space-time, which is introduced to distinguish from the so-called tetrad denoted usually by $e_{\mu}^a(x)$ that is required to be a general covariant vector field under the general coordinate transformations in Einstein theory of general relativity. The bi-covariant vector field $\chi_{\mu}^{\;\; a} (x) $ constitutes a basis for the locally flat non-coordinate space-time and becomes a basic gravitational field in the globally flat Minkowski space-time of coordinates.

In general, when extending the global scaling symmetry of quantum fields to be a local scaling gauge symmetry, but keeping a global scaling symmetry of coordinates, we  shall introduce the Weyl gauge field\cite{WG} $W_{\mu}(x) $ which governs a basic force of {\it scaling gauge interaction} and transforms under the local scaling gauge transformation as follows 
\begin{eqnarray}
 W_{\mu}(x) \to W'_{\mu}(x) = W_{\mu}(x) + g_w^{-1} \partial_{\mu} \ln \xi(x)\, .
\end{eqnarray}
As the local scaling transformation is an Abelian gauge symmetry, the field strength is simply given by 
\be
{\cal W}_{\mu\nu} = \partial_{\mu}W_{\nu} - \partial_{\nu}W_{\mu} \, .
\ee

\section{ Theory of Quantum Gravity with postulates of Gauge Invariance and Coordinate Independence} 

As the bi-covariant vector field $\chi^a_{\;\; \mu}(x)$ or its inverse $\hat{\chi}_a^{\;\;\mu}(x)$ is an essential ingredient for introducing both the local spinnic and scaling gauge transformations of fermion fields and the  global Lorentz and scaling transformations of coordinates. It will be shown that the bi-covariant vector field $\chi^a_{\;\; \mu}(x)$ is a basic field in characterizing the gravitational interactions.  

\subsection{Gravifield Space-time as a Locally Flat Non-coordinate Space-time}

The bi-covariant vector field $\hat{\chi}_a^{\;\;\mu}(x)$ allows us to define a non-coordinate basis in terms of the basis of coordinate space-time. Let us consider the following vector field valued in the Dirac $\gamma$-matrix basis $\{\gamma^a/2\}$
\be
& & \frac{1}{2} \gamma^a\, \hat{\chi}^{\;\;\mu}_a(x)\partial_{\mu} = \frac{1}{2} \gamma^a\, \hat{\chi}_a  = \hat{\chi}^{\;\mu}\partial_{\mu} \; ,
\ee
with
\be
& & \hat{\chi}_a =  \hat{\chi}^{\;\;\mu}_a(x)\partial_{\mu} \, , \quad \hat{\chi}^{\;\mu} = \frac{1}{2} \gamma^a \hat{\chi}^{\;\;\mu}_a(x)\, ,
\ee
where the derivative operator $\partial_{\mu} \equiv \partial/\partial x^{\mu}$ is known to consist of a coordinate basis $\{\partial_{\mu}\}\equiv \{\partial/\partial x^{\mu}\} $ in the globally flat Minkowski space-time.  Thus the vector field $ \hat{\chi}_a$ forms a basis $ \{\hat{\chi}_a \} $ for the locally flat non-coordinate space-time. 

In the coordinate basis, a dual basis $\{dx^{\mu}\}$ is defined as
\be
 < dx^{\nu},\, \partial/\partial x^{\mu}  > = \frac{\partial x^{\nu}}{\partial x^{\mu}} = \eta_{\mu}^{\; \nu} \, .
\ee
The bi-covariant vector field $\chi_{\mu}^{\;\; a}(x)$ dual to $\hat{\chi}_a^{\;\;\mu}(x)$ enables us to define a dual non-coordinate basis in terms of the dual coordinate basis $\{dx^{\mu}\}$.  Consider the following vector field with valued in the Dirac $\gamma$-matrix basis $\{\gamma^a/2\}$
\be
& & \frac{1}{2} \gamma_a\, \chi_{\mu}^{\;\; a} (x) dx^{\mu}  =  \frac{1}{2} \gamma_a\, \chi^{\; a} (x) =  \chi_{\mu}(x) dx^{\mu}\; , \nonumber \\
& & \chi^{\; a}   = \chi_{\mu}^{\;\; a} (x) dx^{\mu} \; , \quad  \chi_{\mu} = \chi_{\mu}^{\;\; a} (x) \frac{1}{2} \gamma_a \, .
\ee
Thus the vector field $\chi^{\, a}$ defines a dual basis $\{\chi^{\, a}\}$ for the locally flat non-coordinate space-time, 
 \be
 < \chi^{\, b},\,  \hat{\chi}_a > = \chi_{\nu}^{\;\; b}(x)  \hat{\chi}_a^{\;\; \mu} (x)  < dx^{\nu} ,\, \partial_{\mu}> =  \chi_{\nu}^{\;\; b}(x)  \hat{\chi}_a^{\;\; \mu} (x)  \eta_{\mu}^{\; \nu} = \eta_a^{\;\, b} \, .
\ee

It is easily shown that the non-coordinate basis ${\hat{\chi}_a }$ does not commute, its commutation relation is given by 
\be
[ \hat{\chi}_a ,\; \hat{\chi}_b] = \chi_{ab}^c\,  \hat{\chi}_c, \qquad  \chi_{ab}^c \equiv - \hat{\chi}_a^{\;\; \mu} \hat{\chi}_b^{\;\; \nu} \chi_{\mu\nu}^c \, ; \qquad \chi_{\mu\nu}^c  = \partial_{\mu}\chi_{\nu}^{\;\; c} - \partial_{\nu}\chi_{\mu}^{\;\; c} 
\ee 
which indicates that the locally flat non-coordinate space-time is in general associated with a non-commutative geometry. Where the gauge-type field tensor $\chi_{\mu\nu}^c $ shall reflect the gravitational field strength.

It is then interesting to regard $\chi^{\, a}(x) $ as an one-form vector in the flat Minkowski space-time, which motivates us to define a gauge-type potential field of gravity 
\be
{\mathsf G}_{\mu}(x) \equiv \chi_{\mu}^{\;\; a} \frac{1}{2} \gamma_a \, , \qquad  {\mathsf G} = -i {\mathsf G}_{\mu} dx^{\mu} \, .
\ee 
We may refer such a gravitational bi-covariant vector field  $\chi^{\;\; a}_{\mu}(x)$ or $\hat{\chi}_{\mu}^{\;\; a}(x)$ as {\it Gravifield} for short.
The non-coordinate bases $\{\chi^{a}\}$ and $\{\hat{\chi}_{a}\}$ are called as {\it Gravifield bases}, and the locally flat non-coordinate space-time spanned by the Gravifield basis $\{\chi^a (x)\} $ is mentioned as {\it Gravifield space-time}. 

\subsection{Gauge Theory of Quantum Gravity with Coordinate Independence }

In the locally flat Gravifield space-time, let us introduce a coordinate-independent exterior differential operator in terms of the Gravifield bases $\chi^{a}$ and $\hat{\chi}_{a}$,
\be
d_{\chi} =  \chi^{a}\wedge \hat{\chi}_a  \, ,   
\ee
which allows us to define coordinate-independent exterior differential forms in the locally flat Gravifield space-time. Thus, the  gauge potential and field strength can be expressed as the one-form and two-form,
\be 
& & \Om = -i \Om_a\, \chi^a\; , \qquad {\cal R} = d_{\chi} \, \Om + \Om\wedge \Om =\frac{1}{2i} {\cal R}_{ab}\, \chi^a\wedge \chi^b \; . 
\ee
Here $\Om_a$ and ${\cal R}_{ab}$ are the spinnic gauge potential and field strength sided on the locally flat Gravifield space-time. Transmuting to the coordinate basis,  $\Om_a$ and ${\cal R}_{ab}$ are correlated to the spinnic gauge potential $\Om_{\mu}$ and field strength ${\cal R}_{\mu\nu}$  in the flat Minkowski space-time as follows
\be
 \Om_a = \hat{\chi}_a^{\;\; \mu} \Om_{\mu} \; , \qquad {\cal R}_{ab} =  \hat{\chi}_a^{\;\; \mu}\,\hat{\chi}_b^{\;\; \nu} {\cal R}_{\mu\nu} \, .
\ee
Similarly,  we have for the internal gauge field and Weyl gauge field,
\be 
& & {\cal A} = -i {\cal A}_a\, \chi^a\; , \qquad {\cal F} = d_{\chi}\, {\cal A} + {\cal A} \wedge {\cal A} = \frac{1}{2i} {\cal F}_{ab}\, \chi^a\wedge \chi^b \; , \nonumber \\
& & W = -i W_a\, \chi^a\; , \qquad {\cal W} = d_{\chi}\, W = \frac{1}{2i} {\cal W}_{ab}\, \chi^a\wedge \chi^b \; .
\ee

For the Gravifield, the gauge potential and gauge covariant field strength in the Gravifield space-time are given by 
\be
{\mathsf G} = -i {\mathsf G}_a \, \chi^a\; , \qquad {\cal G} = d_{\chi} \, {\mathsf G} + \Om\wedge {\mathsf G}  + g_w W \wedge {\mathsf G} = \frac{1}{2i} {\cal G}_{ab}\, \chi^a\wedge \chi^b \; , 
\ee
which relate to the gauge potential ${\mathsf G}_{\mu}$ and field strength ${\cal G}_{\mu\nu}$ in the flat Minkowski space-time as follows
\be
 {\mathsf G}_a = \hat{\chi}_a^{\;\; \mu} {\mathsf G}_{\mu} \; , \qquad {\cal G}_{ab} =  \hat{\chi}_a^{\;\; \mu}\,\hat{\chi}_b^{\;\; \nu} {\cal G}_{\mu\nu} \, ,
\ee
with 
\be
{\cal G}_{\mu\nu} & = & \nabla_{\mu}\chi_{\nu}  - \nabla_{\nu}\chi_{\mu} =  [\, \nabla_{\mu}\chi_{\nu}^{\;\; a} (x)  - \nabla_{\nu}\chi_{\mu}^{\;\; a}(x) \,] \, \frac{1}{2}\gamma_a \equiv {\cal G}_{\mu\nu}^a(x) \, \frac{1}{2}\gamma_a  \; , \nonumber \\
{\cal G}_{\mu\nu}^a(x)  & = & (\partial_{\mu} + g_w W_{\mu} ) \chi_{\nu}^{\;\; a}  + g_s\Om_{\mu\;\; b}^{\;a}  \chi_{\nu}^{\;\; b}  - (\partial_{\nu}+ g_w W_{\nu} )\chi_{\mu}^{\;\; a}   - g_s\Om_{\nu\;\; b}^{\; a}  \chi_{\mu}^{\;\; b} \, .
\ee
The covariant derivative is defined as 
\be
\nabla_{\mu} = \partial_{\mu}  + \Om_{\mu} +  g_w W_{\mu}  = \bigtriangledown_{\mu}  +  g_w W_{\mu} \, , 
\quad \td_{\mu} = \partial_{\mu}  + \Om_{\mu} \, .
\ee

In terms of the exterior differential forms in the locally flat Gravifield space-time, we are able to construct the gauge invariant and coordinate-independent action for the gravitational gauge theory
\begin{eqnarray}
\label{action2}
S_{\chi}  & = & \int \{\, \frac{1}{2} [i \bar{\Psi} \ast \chi \wedge {\mathcal D} \, \Psi\, + i \bar{\psi} \ast \chi \wedge \td \, \psi\, + H.c. ] 
+ y_s Tr\, (\chi \wedge \chi)\wedge \ast (\chi \wedge \chi )\, \bar{\psi} \phi \psi \nonumber \\
& - &  \frac{1}{g_A^2}  Tr\, {\cal F} \wedge \ast {\cal F}   - \frac{1}{g_s^2}  Tr\, {\cal R} \wedge \ast {\cal R}  -    \frac{1}{2} {\cal W} \wedge \ast {\cal W}    +  \frac{1}{2} \, \alpha_W\, \phi^2 Tr\, {\cal G}\wedge \ast {\cal G}  \nonumber  \\
& - &  \frac{1}{2}  d \phi \wedge \ast d \phi   - \alpha_E  Tr\, {\cal R} \wedge \ast (\chi \wedge \chi ) \,  \phi^2  + \lambda_s Tr\, (\chi \wedge \chi)\wedge \ast (\chi \wedge \chi )\, \phi^4 \,  +  {\cal L}' \} \, ,
\end{eqnarray}
The couplings $y_s$, $g_s$, $\alpha_W$, $\alpha_E$ and $\lambda_s$ are the constant parameters. ${\cal L}'$ denotes the Lagrangian density for possible other interactions. 

The above action for the gravitational gauge theory is presented in the locally flat Gravifield space-time, the fermion fields and gauge fields all belong to the spinor representations and vector representations of the spinnic group SP(1,3), respectively. In the above action, we have used the following definitions 
\be
& &  {\mathcal D} = \chi^a {\mathcal D}_a \; , \quad {\mathcal D}_a  =  \hat{\chi}_a -i {\cal A}_{a} - i \Om_{a} =    \hat{\chi}_a^{\;\;\mu} {\mathcal D}_{\mu} = \hat{\chi}_a^{\;\;\mu} (\, \partial_{\mu} -i {\cal A}_{\mu} -i \Om_{\mu}  \, )  \; 
\ee
Here ${\mathcal D} $ defines a covariant exterior differential operator in the lGravifield space-time. The Hodge star in the Gravifield space-time is defined as 
\be
 \ast \chi^a = \frac{1}{3!} \epsilon^a_{\;\; bcd} \,  \chi^b \wedge \chi^d \wedge \chi^c \; ,  \quad \ast {\cal R} = \frac{1}{4i} \epsilon^{ab}_{\;\;\;\; cd}\,  {\cal R}_{ab}\, \chi^c\wedge \chi^d 
\ee
and
\be
& & (\chi \wedge \chi)  = \chi^a \wedge \chi^b\, \frac{1}{2i} \Sigma_{ab} \; , \qquad \ast (\chi \wedge \chi)  = \frac{1}{2} \epsilon^{ab}_{\;\;\;\; cd}\, \chi^c \wedge \chi^d\, \frac{1}{2i} \Sigma_{ab} \; , \nonumber \\
& & d \phi =  (d_{\chi} - i g_w W )\phi\; , \qquad \ast d \phi =  \frac{1}{3!} \epsilon^{a}_{\;\; bcd} \, \chi^b \wedge \chi^d \wedge \chi^c \,  (\hat{\chi}_{a} - g_w W_{a})\phi \; .
\ee
The totally antisymmetric Levi-Civita tensor $ \epsilon^{abcd}$ ( $\epsilon^{0123} = 1$) satisfies the identities 
\be
& & \epsilon^{abcd}\epsilon_{ab}^{\;\;\;\; c'd'} = -2(\eta^{cc'}\eta^{dd'} - \eta^{cd'}\eta^{dc'} ) \; , \nonumber \\
& & \epsilon^{abcd}\epsilon_{abc}^{\;\;\;\;\;\; d'} = -6\eta^{dd'},\qquad \epsilon_{abcd}\epsilon^{abcd} = -24 \,  .
\ee

Note that a scalar field $\phi(x)$ is introduced to maintain both the global scaling and local scaling symmetries for the gauge-type Gravifield interaction characterized by the coupling constant $\alpha_W$ and the scalar-type spinnic gauge interaction characterized by the coping constant $\alpha_E$. We also introduce a singlet fermion field $\psi(x)$ to couple with the scalar field $\phi(x)$ in order to realize a quantum inflation of the universe.  The scalar field $\phi(x)$ characterizes the conformal scaling property, it transforms under the local scaling gauge transformation and the global scaling transformation as 
\begin{eqnarray}
& & \phi(x) \to \phi'(x) = \xi(x) \phi(x) \; , \nonumber \\
& & \phi(x) \to \phi'(x') = \lambda\, \phi(x)  \, , \quad  x^{\mu} \to x^{'\mu} = \lambda^{-1}\, x^{\nu} 
\end{eqnarray}

\section{Quantum Field Theory of Gravity and Conservation Laws}

To make the field theory description on quantum gravity, it is helpful to transmute the above gauge invariant and coordinate-independent action of quantum gravity formulated in the Gravifield space-time into a standard form expressed within the framework of QFT in the globally flat Minkowski space-time. By converting the Gravifield basis into the coordinate basis via the Gravifield, we obtain the following gauge invariant theory of gravity in the globally flat Minkowski space-time
\begin{eqnarray}
\label{action3}
S_{\chi}  & = & \int d^{4}x\; \chi\, \frac{1}{2} \{\,  [\, \hat{\chi}^{\mu\nu} ( \bar{\Psi} \chi_{\mu} i {\mathcal D}_{\nu}   \Psi  + \bar{\psi} \chi_{\mu} i \td_{\nu}   \psi  ) + H.c.\, ] - y_s \bar{\psi} \phi \psi   \nonumber \\
& - &  \frac{1}{4}  \hat{\chi}^{\mu\mu'} \hat{\chi}^{\nu\nu'} [\,  {\cal F}^I_{\mu\nu} {\cal F}^{I}_{\mu'\nu'} + {\cal R}_{\mu\nu}^{ab} {\cal R}_{\mu'\nu'ab} + {\cal W}_{\mu\nu} {\cal W}_{\mu'\nu'}  - \alpha_W\,  \phi^2\,  {\cal G}_{\mu\nu}^a {\cal G}_{\mu'\nu' a}  \, ]\nonumber \\
& + &  \frac{1}{2} \hat{\chi}^{\mu\nu} d_{\mu} \phi d_{\nu}\phi -  \alpha_E g_s \phi^2 \hat{\chi}_a^{\;\;\mu} \hat{\chi}_b^{\;\;\nu}  {\cal R}_{\mu\nu}^{ab} - \lambda_s \phi^4 \, + {\cal L}'(x)   \}  \, ,
\end{eqnarray}
which enables us to provide a unified description for all basic forces based on the framework of QFT for gauge symmetries. Where we have defined the tensor field
\be \label{tensor}
& &  \hat{\chi}^{\mu\nu}(x) = \hat{\chi}_{a}^{\;\;\mu}(x) \hat{\chi}_{b}^{\;\;\nu}(x) \eta^{ab}\; .
\ee
Where the Gravifield $\chi_{\mu}^{\;\; a}(x)$ behaves like a gauge-type field, its dynamics is governed by the field strength of antisymmetric tensor ${\cal G}_{\mu\nu}^a(x)$. Unlike the usual gauge fields, the Gravifield $\chi_{\mu}^{\;\; a}(x)$ couples inversely to all kinematic and interaction terms of  quantum fields. Thus the Gravifield $\chi_{\mu}^{\;\; a}(x)$ is thought to be a basic gauge-type field and its field strength ${\cal G}_{\mu\nu}^a(x)$ characterizes the gravitational force. 
 
\subsection{Equations of Motion for Quantum Fields}  
 
When taking the globally flat Minkowski space-time as an inertial frame for a reference, we are able to describe the motions of quantum fields and make a meaningful definition for the momentum and energy. By applying for the least action principle,  we obtain equations of motion for all fields under an infinitesimal variation of the fields
\be \label{EM0}
 \gamma^a \hat{\chi}_a^{\;\; \mu} i ( {\mathcal D} _{\mu} + {\mathsf V} _{\mu} ) \Psi  =   0\; , \quad \gamma^a \hat{\chi}_a^{\;\; \mu} i ( \td _{\mu} + {\mathsf V} _{\mu} ) \psi  =   0  \; .
\ee
for the fermion fields with the definition for a spinnic gauge invariant vector field 
\be
{\mathsf V}_{\mu} (x) & \equiv &  \frac{1}{2} \hat{\chi}\chi_{\mu}^{\;\; c} \td_{\rho} (\chi \hat{\chi}_c^{\;\; \rho})  \, ,
\ee
and 
\be  
& & D_{\nu} (\chi \hat{\chi}^{\mu\mu'} \hat{\chi}^{\nu\nu'}  {\cal F}^{I}_{\mu'\nu'} )  = J^{I\,\mu} \, ,   \label{EM1} \\
& & \btd_{\nu} ( \chi   \hat{\chi}^{\mu\mu'} \hat{\chi}^{\nu\nu'} {\cal R}_{\mu'\nu' }^{\;\; ab} ) =    J^{\mu\, ab}\, ,   \label{EM2}
\ee
for the internal gauge fields ${\cal A}_{\mu}^I$ and spinnic gauge field $\Om_{\mu}^{ab}$. Where the fermonic vector currents and vector-tensor currents are given by 
\be
 J^{I\,\mu} & = &  g_A \chi  \bar{\Psi} \gamma^a \hat{\chi}_a^{\;\;\mu} T^I \Psi \, ,  \\
 J^{\mu\, ab}  & = & \frac{1}{2} g_s \chi \bar{\Psi} \hat{\chi}_c^{\;\; \mu} \{ \gamma^c\;\;  \frac{1}{2} \Sigma^{ab} \} \Psi  
 + \frac{1}{2} g_s \chi \bar{\psi} \hat{\chi}_c^{\;\; \mu} \{ \gamma^c\;\;  \frac{1}{2} \Sigma^{ab} \} \psi \nonumber \\
& - &  \alpha_E g_s \btd_{\nu} (\chi   \hat{\chi}^{\mu\mu'} \hat{\chi}^{\nu\nu'}   \phi^2 \chi_{\mu'\nu'}^{[ab]} ) +  \frac{1}{2} \chi \phi^2 \alpha_W \hat{\chi}^{\mu \mu'} \hat{\chi}^{\nu\nu'} 
 \chi_{\nu}^{\;\; [a} {\cal G}_{\mu'\nu'}^{b]}  \, ,
\ee
with the notations
\be
& & \chi_{\mu'\nu'}^{[ab]}   =   \chi_{\mu'}^{\;\; a} \chi_{\nu' }^{\;\; b}  - \chi_{\mu' }^{\;\; b} \chi_{\nu' }^{\;\; a}  \, ; \quad 
  \chi_{\nu}^{\;\; [a} {\cal G}_{\mu'\nu'}^{\;\; b]} = \chi_{\nu}^{\;\; a} {\cal G}_{\mu'\nu'}^{\;\; b }  - \chi_{\nu}^{\;\; b}  {\cal G}_{\mu'\nu' }^{\;\; a} \, .
\ee
For the scaling gauge field $W_{\mu}$ and scalar field $\phi$, we obtain the following equations of motion
\be   
& & \partial_{\nu} (\chi   \hat{\chi}^{\mu\mu'} \hat{\chi}^{\nu\nu'}  {\cal W}_{\mu'\nu'})  =   J^{\mu} \, ,   \label{EM3} \\
& &  (\partial_{\mu} + g_w W_{\mu} ) ( \chi \hat{\chi}^{\mu\nu}  d_{\nu}\phi ) =  J \, ,  \label{EM4}
\ee
with the bosonic vector current and scalar current 
\be
J^{\mu} & = &  - g_w\chi \hat{\chi}^{\mu\nu} \phi d_{\nu} \phi 
- g_w   \chi \phi^2  \alpha_W \hat{\chi}^{\mu \mu'} \hat{\chi}^{\nu\nu'} \chi_{\nu}^{\;\; a}  {\cal G}_{\mu'\nu' a}\, , \\
J & = & -\chi y_s\bar{\psi} \psi + \chi \phi [\,  \alpha_W\, \hat{\chi}^{\mu\mu'}  \hat{\chi}^{\nu\nu'}  {\cal G}_{\mu\nu}^a {\cal G}_{\mu'\nu' a}- 2\alpha_E g_s \hat{\chi}_{a}^{\;\;\mu} \hat{\chi}_{b}^{\;\;\nu}  {\cal R}_{\mu\nu}^{ab} - 4\lambda_s \phi^2\, ]\, .
\ee

The equation of motion for the Gravifield $\chi_{\mu}^{\;\; a}$ is given by
\be  \label{EM5}
& & \alpha_W (\td_{\nu} -g_w W_{\nu}) (\, \phi^2  \chi   \hat{\chi}^{\mu \mu'} \hat{\chi}^{\nu\nu'}
  {\cal G}_{\mu'\nu' a} \, ) = J_a^{\;\; \mu} \, ,
\ee
with the bi-covariant vector currents
\be \label{EM4}
 J_a^{\;\; \mu} & = & - \chi \hat{\chi}_a^{\;\;\mu} {\cal L} + \frac{1}{2} \chi \hat{\chi}_a^{\rho}  \hat{\chi}_c^{\;\; \mu}  [ \bar{\Psi} \gamma^c i {\mathcal D}_{\rho} \Psi  + \bar{\psi} \gamma^c i \td_{\rho} \psi   + H.c.\, ] \nonumber \\
& - & \chi \hat{\chi}_a^{\;\; \rho} \hat{\chi}^{\mu \mu'} \hat{\chi}^{\nu\nu'}  [\,  {\cal F}^I_{\rho\nu} {\cal F}^{I}_{\mu'\nu'} + {\cal R}_{\rho\nu}^{cd} {\cal R}_{\mu'\nu'\; cd} +  {\cal W}_{\rho\nu} {\cal W}_{\mu'\nu'}  - 
\alpha_W  \phi^2 {\cal G}_{\rho\nu}^{b} {\cal G}_{\mu'\nu' b}    \, ] \nonumber \\
& + &  \chi \hat{\chi}_a^{\;\; \nu'} \hat{\chi}^{\mu \mu'} d_{\mu'}\phi  d_{\nu'} \phi   - 2 \alpha_E g_s \chi \phi^2  \hat{\chi}_{c}^{\;\;\mu} \hat{\chi}_{a}^{\;\; \mu'}    {\cal R}_{\mu'\nu'}^{cd}\hat{\chi}_{d}^{\;\; \nu'} 
  \, .
\ee

\subsection{Conservation Laws and Extension to Einstein's Gravity Equation  }

With the above QFT description on the gauge theory of gravity in the flat Minkowski space-time, it enables us to present various basic conservation laws. 

\subsubsection{Conservation Laws for Gauge Invariance}   

It is known that within the framework of QFT,  each symmetry corresponds to a conservation law.  For the internal gauge symmetry, from the equation of motion for the internal gauge field ${\cal A}_{\mu}^I$ in Eq.(\ref{EM1}), we can prove that the conservation law still holds  
\be
D_{\mu}J^{I\mu} & = & D_{\mu} D_{\nu} (\chi \hat{\chi}^{\mu\mu'} \hat{\chi}^{\nu\nu'}  {\cal F}^{I}_{\mu'\nu'} ) = 0 \, , 
\ee
where the symmetric property of the tensor $ \hat{\chi}^{\mu\nu} =  \hat{\chi}^{\nu\mu}$ and antisymmetric property of the tensor $ {\cal F}^{I}_{\mu\nu}  $  are only used, which shows that the gravitational effects are totally eliminated. Actually, from the definitions of the fermion vector currents and the covariant derivative, we can also verify the conservation law 
\be
D_{\mu}J^{I\mu} & \equiv &  (\delta^{IK} \partial_{\mu} + f^{IJK} {\cal A}_{\mu} ) (g_A \chi  \bar{\Psi} \gamma^a \hat{\chi}_a^{\;\;\mu} T^K \Psi  )\nonumber \\
& = & g_A \chi \hat{\chi}_a^{\;\;\mu} {\mathcal D}_{\mu} (\bar{\Psi} \gamma^a  T^I \Psi ) + g_A \btd_{\mu} (\chi \hat{\chi}_a^{\;\;\mu} ) \bar{\Psi} \gamma^a  T^I \Psi =0\, ,
\ee
where the equation of motion for the fermion field in Eq.(\ref{EM0}) is used.

Similarly, we can show from the equations of motion in Eqs.(\ref{EM2}) and (\ref{EM3}) the conservation laws for the spinnic and scaling gauge symmetries, 
\be   
\btd_{\mu}J^{\mu\, ab}  & = & \btd_{\mu} \btd_{\nu} (\chi   \hat{\chi}^{\mu\mu'} \hat{\chi}^{\nu\nu'}  {\cal R}_{\mu'\nu' }^{\;\; ab} ) =0  \, , \label{EM2-2} \\
\partial_{\mu} J^{\mu}  & = & \partial_{\mu} \partial_{\nu} (\chi   \hat{\chi}^{\mu\mu'} \hat{\chi}^{\nu\nu'}  {\cal W}_{\mu'\nu'}) \equiv 0 \, \label{EM3-3} .
\ee
From the definitions of the tensor-type current $J^{\mu\, ab}$ and vector current $J^{\mu}$, we have 
\be 
\btd_{\mu}J^{\mu}_{\;\; ab} &  =  & \frac{1}{2} \btd_{\mu} \Sm^{\mu}_{\;\; ab}   + \frac{1}{2} J_{[ab]} - \alpha_E g_s \phi^2 \chi  (  \hat{\chi}_a^{\;\;\mu} {\cal R}_{\mu\nu b}^{\;\;\;\;\;\, c}   -  \hat{\chi}_b^{\;\;\mu} {\cal R}_{\mu\nu a}^{\;\;\;\;\;\, c}  )\hat{\chi}_c^{\;\;\nu} =0 \, ,  \label{CLS} \\
\partial_{\mu} J^{\mu} & = & \partial_{\mu} ( \chi \hat{\chi}^{\mu\nu} \phi d_{\nu} \phi 
+ \alpha_W \chi \phi^2  \hat{\chi}^{\mu \mu'} \hat{\chi}^{\nu\nu'} \chi_{\nu}^{\;\; a}  {\cal G}_{\mu'\nu' a} )= 0 \, , \label{CLSC}
\ee
with the definitions
\be 
& & \Sm^{\mu}_{\;\; ab} = g_s \chi [\,  \bar{\Psi} \hat{\chi}_c^{\;\; \mu} \{ \gamma^c\;\;  \frac{1}{2} \Sigma_{ab} \} \Psi  + \bar{\psi} \hat{\chi}_c^{\;\; \mu} \{ \gamma^c\;\;  \frac{1}{2} \Sigma_{ab} \} \psi \,] \; , \nonumber \\
& & J_{[ab]} = J_a^{\mu} \chi_{\mu b} - J_b^{\mu}\chi_{\mu a} \, .
\ee
where the tensor $\Sm^{\mu}_{\;\; ab} $ corresponds to the spin angular momentum tensor and $J_{[ab]}$ is related to the energy-momentum tensor. It is seen that the tensor current $J^{\mu\, ab}$ and the Gravifield current $J_a^{\; \mu}$ are in general correlated,  which is easily understood as the spinnic gauge field $\Om_{\mu}^{ab}$ and Gravifield $\chi_{\mu}^{\;\; a}$ have to be introduced simultaneously when gauging the spinnic symmetry of fermions. 

We now come to the conclusion that in the presence of gravity the conservation laws for the various currents due to gauge symmetries hold when the gravitational effects are included.

\subsubsection{Energy-Momentum Conservation}

As the above theory of quantum gravity is built based on the framework of relativistic QFT in the flat Minkowski space-time,  the differences of the spacial coordinates or time coordinates can be defined by the standard ways proposed in the special relativity, which allows us to make a meaningful definition for the energy-momentum and angular momentum as well as their conservation laws.

The translational symmetry of the theory leads to the energy-momentum conservation law. Following the standard method, let us consider an infinitesimal translational transformation of coordinates $x^{\mu} \to x^{'\mu} = x^{\mu} + a^{\mu}$, the variation of action is given by 
\be 
\delta S_{\chi} = \int d^4x\, \partial_{\mu} ( {\cal T}^{\;\, \mu}_{\nu}) a^{\nu} = 0 \, , \nonumber 
\ee
where the surface term is ignored. Thus for an arbitrary displacement $a^{\nu}$, it leads to the well-known energy-momentum conservation law
\begin{equation} \label{EMC}
\partial_{\mu} {\cal T}^{\;\, \mu}_{\nu} = 0 \, ,
\end{equation}
with the gauge invariant energy-momentum tensor given by
\be \label{EMT}
{\cal T}^{\;\, \mu}_{\nu} & = & - \eta^{\mu}_{\; \nu}\chi {\cal L}  + \frac{1}{2}\chi \hat{\chi}^{\;\; \mu}_{a} [\, 
i\bar{\Psi} \gamma^{a}{\mathcal D}_{\nu}\Psi + i\bar{\psi} \gamma^{a} \td_{\nu}\psi  + H.c.\,  ] \nonumber \\
& - & \chi \hat{\chi}^{\mu\mu'} \hat{\chi}^{\rho\sigma}  [\,  {\cal F}^I_{\mu'\rho} {\cal F}^{I}_{\nu \sigma} 
+  {\cal R}_{\mu'\rho}^{ab}  {\cal R}_{\nu \sigma\, ab} + {\cal W}_{\mu'\rho} {\cal W}_{\nu \sigma} 
-  \alpha_W \phi^2 {\cal G}_{\mu'\rho}^a {\cal G}_{\nu \sigma a}   \, ] \nonumber \\
& + & \chi \hat{\chi}^{\mu\mu'} d_{\mu'} \phi d_{\nu}\phi  - 2 \alpha_E g_s\chi \phi^2 \hat{\chi}_{a}^{\;\;\mu}  {\cal R}_{\nu \rho}^{ab} \hat{\chi}_{b}^{\;\;\rho} \, , 
\ee
which is the total energy-momentum tensor from all fields including gravitational fields. It is intriguingly noted that the gauge invariant energy-momentum tensor ${\cal T}_{\mu\nu}$ is in general not symmetric even when freezing out the gravitational interactions due to the existence of fermionic fields 
\be 
{\cal T}_{\mu\nu} \neq {\cal T}_{\nu\mu} \, .
\ee

\subsubsection{ Conservation Laws for Global Lorentz and Scaling Invariances}

Similar to the energy-momentum conservation of translational invariance,  considering an infinitesimal transformation $x'^{\mu} = x^{\mu} + \delta L^{\mu}_{\; \nu} x^{\nu}$, we arrive at the conservation law for the global Lorentz transformation invariance
\be
\partial_{\mu} \Lm^{\mu}_{\;\, \rho\sigma}- {\mathsf T}_{[\rho\sigma]} =0 \, ,
\ee
 with
 \be \label{AMC}
  \Lm^{\mu}_{\;\, \rho\sigma} \equiv {\cal T}^{\;\,\mu}_{\rho}\, x_{\sigma} - {\cal T}^{\;\, \mu}_{\sigma}\, x_{\rho}  \, , \qquad {\mathsf T}_{[\rho\sigma]} \equiv {\cal T}_{\rho}^{\;\,\rho'} \eta_{\rho'\sigma} - {\cal T}_{\sigma}^{\;\,\sigma'}\eta_{\sigma'\rho}  \, .
 \ee
Here $\Lm^{\mu}_{\;\, \rho\sigma}$ defines an orbital angular momentum tensor of space-time. It is shown that the orbital angular momentum tensor is in general not conserved homogeneously, i.e.,  $\partial_{\mu} \Lm^{\mu}_{\;\, \rho\sigma}\neq 0$, 
due to the asymmetric energy-momentum tensor ${\mathsf T}_{\rho\sigma} \neq {\mathsf T}_{\sigma\rho}  $.
 
 From the conservation law Eq.(\ref{CLS}) for the spinnic gauge invariance, we have
\be \label{spinningC}
 & & \partial_{\mu} \Sm^{\mu}_{\;\; \rho\sigma}  + {\cal  T}_{[\rho\sigma]} - \Sm^{\mu}_{\;\; ab} (\nabla_{\mu}\chi_{\rho}^{\;\, a}\,  \chi_{\sigma}^{\;\, b}  + \chi_{\rho}^{\;\, a}\,  \nabla_{\mu} \chi_{\sigma}^{\;\, b}  ) \nonumber \\
 & & - 2\alpha_E g_s \phi^2 \chi  (  \hat{\chi}_a^{\;\,\mu} {\cal R}_{\mu\nu b}^{\;\;\;\;\;\,c} - \hat{\chi}_b^{\;\,\mu} {\cal R}_{\mu\nu a}^{\;\;\;\;\;\,c}  )\hat{\chi}_c^{\;\,\nu} \chi_{\rho}^{\;\, a} \chi_{\sigma}^{\;\, b} =0 \, ,
\ee
where $\Sm^{\mu}_{\;\; \rho\sigma} $ defines the spinnic angular momentum tensor with ${\cal T}_{[\rho\sigma]}$ relating to the energy-momentum tensor
\be 
\Sm^{\mu}_{\;\; \rho\sigma}  & = & \Sm^{\mu}_{\;\; ab} \chi_{\rho}^{\;\, a} \chi_{\sigma}^{\;\, b} = g_s \chi \hat{\chi}_c^{\;\; \mu} [\, \bar{\Psi}  \{ \gamma^c\;\;  \frac{1}{2} \Sigma_{ab} \} \Psi  + \bar{\psi}  \{ \gamma^c\;\;  \frac{1}{2} \Sigma_{ab} \} \psi\, ]  \chi_{\rho}^{\;\, a} \chi_{\sigma}^{\;\, b}  \, , \nonumber \\
{\cal T}_{[\rho\sigma]}  & = & J_{[ab]} \chi_{\rho}^{\;\, a} \chi_{\sigma}^{\;\, b} = {\cal T}_{\rho}^{\;\; \mu} \chi_{\mu\sigma}  - {\cal T}_{\sigma}^{\;\; \mu} \chi_{\mu\rho}   \, , \quad \chi_{\mu\nu} = \chi_{\mu}^{\;\; a}\chi_{\nu}^{\;\; b} \eta_{ab} \, .
\ee 
where $\chi_{\mu\nu} $ is dual to the symmetric tensor field $\hat{\chi}^{\mu\nu}$ in Eq.(\ref{tensor}).
 
It is interesting to introduce a total angular momentum tensor 
\be
 {\cal J}^{\mu}_{\;\; \rho\sigma}  & \equiv & \Lm^{\mu}_{\;\; \rho\sigma}  + \Sm^{\mu}_{\;\; \rho\sigma}  \, ,
\ee
we obtain a new form of conservation law from Eq.(\ref{AMC}) and Eq.(\ref{spinningC})
\be 
 & & \partial_{\mu} {\cal J}^{\mu}_{\;\; \rho\sigma}  -({\mathsf T}_{[\rho\sigma]} -  {\cal T}_{[\rho\sigma]} ) - \Sm^{\mu}_{\;\; ab} (\nabla_{\mu}\chi_{\rho}^{\;\, a}\,  \chi_{\sigma}^{\;\, b}  + \chi_{\rho}^{\;\, a}\,  \nabla_{\mu} \chi_{\sigma}^{\;\, b}  ) 
 \nonumber \\
 & &  -   2\alpha_E g_s \phi^2 \chi  (  \hat{\chi}_a^{\;\,\mu} {\cal R}_{\mu\nu b}^{\;\;\;\;\;\,c}   -  \hat{\chi}_b^{\;\,\mu} {\cal R}_{\mu\nu a}^{\;\;\;\;\;\,c}  )\hat{\chi}_c^{\;\,\nu} \chi_{\rho}^{\;\, a} \chi_{\sigma}^{\;\, b} = 0 \, .
\ee
When turning the spinnic and scaling gauge symmetries into global Lorentz and scaling symmetries, namely 
\be
 & & \Om_{\mu}^{ab} \to 0\, , \quad \hat{\chi}_a^{\;\;\mu} \to \eta_a^{\;\; \mu}\, , \quad W_{\mu} \to 0 \, , 
 \ee
 we arrive at the familiar conservation law for a total angular momentum tensor
 \be
\partial_{\mu} {\cal J}^{\mu}_{\;\; \rho\sigma}  & = & \partial_{\mu} ( \Lm^{\mu}_{\;\; \rho\sigma}  + \Sm^{\mu}_{\;\; \rho\sigma}  )  =0 \, ,
\ee
which reproduces the result for the gravity-free theory in the Minkowski space-time. This is due to the fact that
 \be
 & &  \partial_{\mu} \Sm^{\mu}_{\;\; \rho\sigma}  = -  {\mathsf T}_{[\rho\sigma]}\, , \quad  
 \partial_{\mu} \Lm^{\mu}_{\;\; \rho\sigma}  = {\mathsf T}_{[\rho\sigma]} \, ,
 \ee
which shows that neither the angular momentum tensor nor the spinnic momentum tensor is conserved homogeneously due to the asymmetric part of the energy-momentum tensor in the presence of fermion fields, only the total angular momentum becomes homogeneously conserved via the cancelation in their asymmetric part of the energy-momentum tensor.

The global scaling symmetry leads to the following conservation law 
\be
(x^{\mu} \frac{\partial}{\partial x^{\mu}} + 4) (\chi\, {\cal L} ) + \partial_{\mu} {\cal T}^{\mu}  - T = 0 \, ,
\ee
 with $ {\cal T}^{\mu} $ and ${\cal T}$ relating to the energy-momentum tensor
 \be \label{SCL}
  {\cal T}^{\mu} \equiv {\cal T}^{\;\, \mu}_{\nu}\, x^{\nu}  \, , \qquad {\cal T} \equiv {\cal T}_{\nu}^{\;\, \mu} \eta_{\mu}^{\;\, \nu}
 = {\cal T}_{\mu}^{\; \mu}  \, .
 \ee
 Notice the fact that the integral $\int d^4x \lambda^4\,  \chi(\lambda x)\, {\cal L}(\lambda x) $ is independent of $\lambda$, the differentiation with respect to $\lambda$ at $\lambda =1$ gives an identity
 \be
 \int d^4 x (x^{\mu} \frac{\partial}{\partial x^{\mu}} + 4) (\chi {\cal L} ) = 0 \, ,\nonumber 
 \ee  
the conservation law is given by
 \be
 \partial_{\mu} {\cal T}^{\mu}  - {\cal T} = 0 \, .
 \ee
Only when the scalar field is freezed out in the gravity-free theory,  the energy-momentum tensor becomes traceless and a homogeneous conservation law for the scaling invariance is reached
\be
\partial_{\mu} {\cal T}^{\mu} = 0\, . 
 \ee

 \subsubsection{Gravity Equation and Conservation Law for Gravifield Tensor}

For the conservation of the bi-covariant vector current $J_a^{\;\; \mu}$, from the equation of motion for the Gravifield $\chi_{\mu}^{\;\; a}$ Eq.(\ref{EM5}), the gauge and Lorentz covariant derivative to the current is given by  
\be  \label{EM3-3}
& & (\td_{\mu} -g_w W_{\mu}) J_a^{\;\; \mu} = \alpha_W (\td_{\mu} -g_w W_{\mu})  (\td_{\nu} -g_w W_{\nu})  (\,  \phi^2  \chi  \hat{\chi}^{\mu \mu'} \hat{\chi}^{\nu\nu'} {\cal G}_{\mu'\nu' a}\,  )   \, \nonumber  \\
& &  = \frac{1}{2}  \alpha_W  \phi^2  \chi  ( g_s {\cal R}_{\mu\nu a }^{\;\;\;\;\; \; b}  - g_w {\cal W}_{\mu\nu} \eta_{a}^{\;\; b}   )  \hat{\chi}^{\mu \mu'} \hat{\chi}^{\nu\nu'}  {\cal G}_{\mu'\nu' b} \, ,
\ee
which is not conserved homogeneously. This is understandable from the fact that the Gravifield behaves not like the usual gauge field as it is introduced to be an accompaniment of spinnic and scaling gauge fields to maintain the spinnic and scaling gauge symmetries. 
 
It is interesting to observe the following relation between the energy-momentum tensor and the bi-covariant vector current for the Gravifield 
\be
 {\cal T}^{\;\, \mu}_{\nu} & = & \chi_{\nu}^{\;\; a} J^{\; \mu}_a  \, ,
\ee
which motivates us to derive the equation of motion for the Gravifield in connection directly with the energy-momentum tensor
\be \label{GE2}
 \partial_{\rho} {\cal G}^{\;\mu\rho}_{ \nu}  - {\cal G}_{\nu}^{\;\; \mu} = {\cal T}^{\;\, \mu}_{\nu} \, , 
\ee
which is referred as the {\it Gravity Equation} alternative to the Einstein's equation for the theory of general relativity. Where we have introduced  the definitions
\be
& & {\cal G}^{\; \mu\rho}_{ \nu}  \equiv \alpha_W \phi^2  \chi  \hat{\chi}^{\mu \mu'} \hat{\chi}^{\rho\nu'}  \chi_{\nu}^{\;\; a}\, {\cal G}_{\mu'\nu' a}  = - {\cal G}^{\; \rho\mu}_{ \nu} \, , \nonumber \\
& &   {\cal G}_{\nu}^{\;\; \mu} \equiv  \alpha_W \phi^2  \chi   \hat{\chi}^{\mu \mu'} \hat{\chi}^{\rho\nu'} ( \nabla_{\rho} \chi_{\nu}^{\;\; a} ) \, {\cal G}_{\mu'\nu' a}  =  (\hat{\chi}_a^{\; \sigma}\nabla_{\rho} \chi_{\nu}^{\;\; a} )\,
{\cal G}^{\; \mu\rho}_{ \sigma}   \, .
 \ee 
Here ${\cal G}^{\; \mu\rho}_{ \nu}$ is mentioned as the gauge invariant {\it Gravifield tensor} and $ {\cal G}_{\nu}^{\;\; \mu} $ as the gauge invariant {\it Gravifield tensor current}. 

From the energy-momentum conservation $\partial_{\mu} {\cal T}^{\;\, \mu}_{\rho}  =  \partial_{\mu} ( J^{\;\mu}_a \chi_{\rho}^{\;\; a} )  =0$, we obtain the conserved current
\be
\partial_{\mu}{\cal G}_{\nu}^{\;\; \mu} =  \partial_{\mu}( \hat{\chi}_a^{\; \sigma}\nabla_{\rho} \chi_{\nu}^{\;\; a} 
{\cal G}^{\; \mu\rho}_{ \sigma} ) =\partial_{\mu}  (\,  \phi^2  \chi  \hat{\chi}^{\mu \mu'} \hat{\chi}^{\rho\nu'} \nabla_{\rho} \chi_{\nu}^{\;\; a}\, {\cal G}_{\mu'\nu' a} \, )=  0 \, ,
 \ee 
which presents an alternative conservation law for the {\it Gravifield tensor current}.


\section{Gauge Symmetry Breaking and Dynamics of Background Fields }

In the above gauge theory of quantum gravity, we have introduced the spinnic gauge field $\Om_{\mu}^{ab}(x)$ and Gravifield $\chi_{\mu}^{\;\; a}(x)$ as well as the scaling gauge field $W_{\mu}(x)$ and the scalar field $\phi(x)$. They are in general massless without considering gauge symmetry breaking.  As there are no experimental indications for those fields,  either they are very heavy or their interactions are very weak. Let us consider possible gravitational gauge symmetry breaking and study the evolution of the universe. 

\subsection{Gauge Symmetry Breaking in Theory of Quantum Gravity}

Before considering gravitational gauge symmetry breaking, we first make two special remarks for the scaling gauge transformations. The scaling gauge symmetry allows us to make a special scaling gauge transformation $\xi_a(x) =1/a_{\chi}(x)$, so that 
\be
\label{GFC}
\phi(x) \to \phi'(x) = \xi_a(x) \phi(x) =  \phi(x)/a_{\chi}(x) = M_S \, .
\ee
Here $M_S$ is regarded as the basic {\it scaling energy scale} to fix the scaling gauge transformation.  Alternatively, we can always choose a special gauge fixing condition, so that the determinant of the Gravifield $\chi = \det \chi_{\mu}^{\;\; a}$ is rescaled into unity
\be
\chi(x) \to \chi'(x) = \xi_{\chi}^{-4}(x) \chi(x) =  1 \,\quad a_{\chi}(x)\to a(x) = \xi_{\chi}(x) a_{\chi}(x)  = \chi^{1/4}(x)a_{\chi}(x) \, . 
\ee
and the scalar field can be expressed as the following  general form
\be
\phi(x) \equiv M_S\, a(x) \, , \quad \chi = \det \chi_{\mu}^{\;\; a} = 1.
\ee 

We now turn to discuss the gravitational gauge symmetry breaking. Let us make a postulate that the gravitational gauge symmetry is broken down in such a way that the theory remains keeping a global Lorentz symmetry.  With this postulate, we consider the following simple background structure under the scaling gauge fixing condition  $\chi(x) =  1$
\be 
& & \langle \chi_{\mu}^{\;\; a}(x) \rangle = \chib_{\mu}^{\;\; a}(x) = \eta_{\mu}^{\;\; a}  \, , \quad \langle \phi(x) \rangle = \bar{\varphi}(x) \equiv \ab(x)\, M_S\, ,  \nonumber \\
& &  \langle \Om_{\mu}^{ab}(x) \rangle =   \eta_{\mu\rho}^{[ab]} \bar{\omega}^{\rho}(x)\, , \quad \eta_{\mu\rho}^{[ab]} \equiv  \eta_{\mu}^{\;\; a} \eta_{\rho}^{\;\;b} - \eta_{\mu}^{\;\; b} \eta_{\rho}^{\;\;a}
 \, , \nonumber \\
& & \langle W_{\mu} \rangle = \bar{w}_{\mu}(x) \, , \quad  \langle \Psi(x) \rangle = 0\, , \quad  \langle \psi(x) \rangle = 0\, ,\quad \langle {\cal A}_{\mu}^I(x) \rangle = 0 \, ,
\ee
with $\ab(x)$, $\ob_{\mu}(x)$ and $\wb_{\mu}(x)$ the gravitational background fields.

\subsection{Dynamics of Background Fields and Cosmic Vector}

To investigate the dynamics of background fields, we shall find out solutions of the background fields $\pb(x)$, $\ob(x)$ and $\wb(x)$ by solving their field equations of motion. 

The field strengths and their scalar products for the background fields are found to be
\be
\bar{R}_{\mu\nu}^{ab} & = &  d_{\mu}\ob^{\rho} \eta_{\nu\rho}^{[ab]} - d_{\nu}\ob^{\rho} \eta_{\mu\rho}^{[ab]} - g_s \ob_{\rho}^2 \eta_{\mu\nu}^{[ab]}    \, , \nonumber \\
\bar{G}_{\mu\nu}^a & = & \nabla_{\mu} \eta_{\nu}^a - \nabla_{\nu}\eta_{\mu}^a = \eta_{\mu}^a \Ob_{\nu} - \eta_{\nu}^a\Ob_{\mu} \, , \quad \Ob_{\mu} \equiv g_s \ob_{\mu} - g_w \wb_{\mu}\, , \nonumber \\
\bar{W}_{\mu\nu} & = & \partial_{\mu} \wb_{\nu} - \partial_{\nu} \wb_{\mu} \, \\
\bar{R} & = & \eta_a^{\mu} \eta_b^{\nu} \bar{R}_{\mu\nu}^{ab}  = - 6 (\partial_{\mu}\ob^{\mu} 
+ g_s \ob_{\mu}\ob^{\mu} ) \, , \nonumber \\
\bar{R}_{\mu\nu}^{ab} \bar{R}^{\mu\nu}_{ab} & = & 4 (d_{\mu}\ob^{\mu} )^2 + 8 d_{\mu}\ob_{\nu}  d^{\mu}\ob^{\nu} + 24 g_s ( d_{\mu}\ob^{\mu} + g_s \ob_{\mu}\ob^{\mu} )\ob_{\nu}\ob^{\nu} \, . \nonumber 
\ee
The corresponding Lagrangian for the background fields is given by
\be
\bar{\cal L} & = &  - (d_{\rho}\ob^{\rho} )^2 -2 d_{\rho}\ob_{\sigma} d^{\rho}\ob^{\sigma} -6g_s \partial_{\rho}\ob^{\rho} \ob_{\sigma}\ob^{\sigma} + \frac{3}{2} \alpha_W  \Ob_{\rho}\Ob^{\rho} \pb^2 \nonumber \\
& & -\frac{1}{4} \bar{W}_{\rho\sigma}\bar{W}^{\rho\sigma} + \frac{1}{2} d_{\rho}\pb d^{\rho}\pb + 6\alpha_E g_s (\partial_{\rho}\ob^{\rho} + g_s \ob_{\rho}\ob^{\rho} )\pb^2 - \lambda_s \pb^4 \, ,
 \ee
with the definitions 
\be \label{CDV}
& & d_{\mu}\pb \equiv (\partial_{\mu} - g_w \wb_{\mu}) \pb \, , \quad d_{\mu}\ob^{\rho}  \equiv (\partial_{\mu} - g_s \ob_{\mu}) \ob^{\rho} \, .
\ee

 The equations of motion for background fields easily read from Eqs.(\ref{EM2})-(\ref{EM5}) as follows,
\be  \label{BEM1}
& & [\, - \partial_{\sigma}\partial^{\sigma} \ob^{\rho} + g_s \ob^{\sigma}\partial_{\sigma} \ob^{\rho} + 2 g_s \ob^{\rho}\partial_{\sigma}\ob^{\sigma} - 3 g_s \ob^{\sigma}\partial^{\rho}\ob_{\sigma} \nonumber \\ 
& + & 2 g_s^2 \ob_{\sigma}\ob^{\sigma}\ob^{\rho} + \frac{1}{2} \alpha_W  g_s \Ob^{\rho} \pb^2 + \alpha_E g_s \partial^{\rho} \pb^2 - 2\alpha_E g_s^2 \ob^{\rho} \pb^2 \, ] \eta_{\mu\rho}^{[ab]} \nonumber \\
& & + [ \,  \partial_{\mu} \partial^{\sigma}\ob^{\rho} - 2g_s \ob_{\mu} \partial^{\sigma}\ob^{\rho} - g_s \ob^{\rho} \partial^{\sigma} \ob_{\mu} - g_s \ob^{\sigma} \partial_{\mu} \ob^{\rho} \, ] \eta_{\sigma\rho}^{[ab]} = 0 \, , 
\ee
for the background spinnic gauge field $\langle \Om_{\mu}^{ab}(x)\rangle$, and 
\be  \label{BEM2}
& & \eta_{\mu}^a \alpha_W (\partial^{\rho} - g_w \wb^{\rho}) (\pb^2\Ob_{\rho}) - \eta_{\rho}^a \alpha_W  [\, (\partial^{\rho} - g_w \wb^{\rho}) (\pb^2\Ob_{\rho}) + g_s (\ob_{\mu} \Ob^{\rho} + 2 \Ob_{\mu}\ob^{\rho} ) \pb^2 \, ] \nonumber \\
& & = - \eta_{\mu}^a 2[\, d_{\rho}\ob_{\sigma} d^{\rho}\ob^{\sigma} + 2 g_s d_{\rho}\ob^{\rho} \ob_{\sigma}\ob^{\sigma} 
+ 3 g_s^2  (\ob_{\sigma}\ob^{\sigma})^2 - \alpha_W \Ob_{\sigma}\Ob^{\sigma}\pb^2 \, ] \nonumber \\
& & \;\;\; - \eta_{\mu}^a 2 \alpha_Eg_s (d_{\rho}\ob^{\rho} + 3g_s \ob_{\rho}\ob^{\rho})  \pb^2 
-\eta_{\rho}^a [\,  \bar{W}_{\rho\sigma}\bar{W}^{\mu\sigma} -  d^{\rho}\pb d_{\mu}\pb - 2 \alpha_W  \Ob^{\rho}\Ob_{\mu} \pb^2 ]\nonumber \\
& &\;\;\; -\eta_{\rho}^a 2[\, d_{\rho}\ob_{\sigma} d_{\mu}\ob^{\sigma}  +  (d_{\rho}\ob_{\mu} + d_{\mu}\ob_{\rho} ) d_{\sigma}\ob^{\sigma} - d_{\sigma}\ob_{\mu} d^{\sigma}\ob^{\rho} + 2 g_s  (d_{\rho}\ob_{\mu} + d_{\mu}\ob_{\rho} ) \ob_{\sigma}\ob^{\sigma} \, ] \nonumber \\
& & \;\;\; + \eta_{\rho}^a  4 \alpha_Eg_s \pb^2 d^{\rho}\ob_{\mu} -\eta_{\mu}^a \bar{\cal L} \, ,
\ee
for the background Gravifield $\langle \chi_{\mu}^{\;\;a}(x)\rangle$, as well as 
\be  
& & (\partial_{\mu} + g_w \wb_{\mu}) d^{\mu} \pb = 6 \alpha_W \Ob_{\mu}\Ob^{\mu} \pb^2 
+ 12 \alpha_Eg_s (\partial_{\mu}\ob^{\mu} + g_s \ob_{\mu}\ob^{\mu})  \pb - 4 \lambda_s \pb^3  \,  ,  \label{BEM3}  \\
& & \partial_{\nu} \bar{W}^{\mu\nu} = - g_w \pb d^{\mu} \pb - 3 \alpha_W  \Ob^{\mu} \pb^2 \, ,   \label{BEM4} 
\ee
 for the background scaling gauge field and scalar field, respectively.

To obtain solutions of background fields, it is helpful to simplify the above equations of motion. For that, we may make a simple and rational {\it ansatz} that the conformally covariant derivative of the background scalar field defined in Eq.(\ref{CDV}) vanishes
\be  \label{BSL}
 d_{\mu} \pb = 0  \, , \quad \mbox{i.e. } \quad g_w\wb_{\mu}(x) = \partial_{\mu} \ln \pb(x) \, . 
\ee                                                                                                                                                                                                                                                                                                                                                                                                                                                                                                                                                                                                                                                                                                                                                            
Such an ansatz indicates that the background scaling gauge field $\wb(x)$ is a pure gauge field and solely determined by the background scalar field $\pb(x)$. 

Applying such a simple ansatz to equation of motion Eq.(\ref{BEM4}) for the background scaling gauge field, we yield the results
\be  \label{BSL0}
& & \bar{W}_{\mu\nu} = 0\, , \quad  \Ob^{\mu}  =0 \, , \quad  \bar{G}_{\mu\nu} = 0,  \nonumber \\
& &  g_s \ob_{\mu}(x) = g_w \wb_{\mu}(x) = \partial_{\mu} \ln \pb(x)\, ,
\ee
which shows that once making a postulate that the background scalar field has a vanishing conformally covariant kinetic energy, the field strength for both the background spinnic gauge field and Gravifield is also vanishing and all the background fields are governed by the background scalar field.  

With the above results, the complicated equations of motion Eqs.(\ref{BEM1})-(\ref{BEM3}) are simplified, respectively, to be
\be
& & 3 (\partial_{\mu}\pb(x) ) \partial_{\nu}^2 \pb(x) = \pb(x) \partial_{\mu} (\partial_{\nu}^2 \pb(x) )\, ,  \label{BSL1} \nonumber  \\
& & 2\partial_{\mu}\pb \partial_{\nu}\pb = \pb \partial_{\mu}\partial_{\nu} \pb \, , \quad \mbox{i.e.} \quad  d_{\mu}\ob_{\nu} = 0  \, ,  \label{BSL2} \nonumber  \\
& & 3\alpha_E \partial_{\nu}^2 \pb(x) = \lambda_s \pb^3(x)\, \label{BSL3}  .
\ee

It is not difficult to solve the above simplified equations. An exact solution for the background scalar field is found to be
\be  \label{BSL4}
\pb(x) = \frac{m_{\kappa} }{\alpha_S(1 - x^{\mu}\kappa_{\mu})}\, , \quad m_{\kappa} = \sqrt{  \kappa_{\mu}\kappa^{\mu} } \, .
\ee
Here $\kappa_{\mu}$ appears to be a constant {\it cosmic vector} with the {\it cosmological mass} scale $m_{\kappa} $. For nonzero constants $\alpha_E$ and $\lambda_s$, they are related to the constant parameter $\alpha_S$ via the following relation
\be  \label{BSL5}
\lambda_s =  6\alpha_E \alpha_S^2\, .
\ee 

As the equations of motion possess a mirror symmetry $Z_2$: $\pb \to - \pb$, and also a reflection symmetry of space-time: $x^{\mu}\to - x^{\mu}$, $\pb(x)\to \pb(-x)$, the solutions must also be invariant under the transformation: $x^{\mu}\to - x^{\mu}$, $\kappa_{\mu}\to - \kappa_{\mu}$. In general, there are four solutions for the background scalar field
\be
\pb(x) \to \pb_{\kappa\pm}^{(\pm)} (x) = \pm \pb_{\kappa\pm} (x) = \pm \frac{ \mk}{\alpha_S(1\mp x^{\mu}\kappa_{\mu})} \, .
\ee

\section{Background Structure and Geometry of Gravifield Space-time and Evolution of Early Universe}

With the above solutions of the background fields, we are able to investigate the background structure of Gravifield Space-time and the properties of {\it background Gravifield Space-time} resulting from the solutions. Based on the {\it background Gravifield Space-time}, it enables us to explore the evolution of early universe. 

\subsection{ Geometry of Gravifield Space-time }

To study the geometry of Gravifield Space-time, we shall define a spinnic and scaling gauge invariant line element with the Gravifield basis $\chi^a$
\be
l^2_{\chi} = a^2_{\chi} \, \eta_{ab}\, \chi^a \chi^b  \, , \quad a_{\chi}\equiv \phi^2/M_S^2 \, .
\ee
Here the multiplying factor $a_{\chi}$ given by the scalar field ensures an invariant line element under both the local and global scaling transformations.  When turning to the coordinate basis in the flat Minkowski space-time, we can express the invariant line element as the following form
\be
l_{\chi}^2  \equiv a_{\chi}^2(x) \chi_{\mu\nu}(x) dx^{\mu} dx^{\nu}  \, , \quad 
 \chi_{\mu\nu}(x) =\chi_{\mu}^{\;\; a}(x) \chi_{\nu}^{\;\; b}(x) \eta_{ab} \, ,
\ee
where $a_{\chi}^2(x)$ is regarded as a {\it conformal scale field} and $\chi_{\mu\nu}(x) $ defines a Lorentz covariant metric tensor field referred as {\it Gravimetric field} for short. Both the conformal scale field and Gravimetric field characterize the geometric properties of Gravifield space-time in such a {\it conformal basis}. 

The scaling gauge invariant line element allows us to choose, by making an appropriate scaling gauge transformation, a convenient basis to investigate the features of Gravifield space-time. For instance, by making a particular scaling gauge transformation,  
\[ a_{\chi} (x) \to a_{\chi}^E (x) = \xi_a(x) a_{\chi}(x) =  1\; , \quad  \chi_{\mu}^{\;\; a}(x) \to \chi_{\mu}^{E\, a}(x) = \xi_a^{-1}(x) \chi_{\mu}^{\;\; a}(x) = a_{\chi}(x) \chi_{\mu}^{\;\; a}(x)\, ,  \]
the line element can be rewritten as
\be
l_{\chi}^2 = a_{\chi}^2(x) \chi_{\mu\nu}(x) dx^{\mu} dx^{\nu}  \equiv  \chi_{\mu\nu}^E (x) dx^{\mu} dx^{\nu}  \, ,\quad 
\chi_{\mu\nu}^E (x) =\chi_{\mu}^{E\, a}(x) \chi_{\nu}^{E\, b}(x) \eta_{ab} \, ,
\ee
which yields an {\it  Einstein-type basis } in the flat Minkowski space-time of coordinates.  

Alternatively, by taking a special scaling gauge transformation, 
\[ a_{\chi}(x)\to a_U(x) = \chi^{1/4}(x)a_{\chi}(x)\, , \quad \chi_{\mu}^{\;\; a}(x) \to \chi_{\mu}^{U\, a}(x) = \chi^{-1/4}(x) \chi_{\mu}^{\;\; a}(x)\, , \]  
we obtain a specific line element
\be
& & l_{\chi}^2  =   a_{\chi}^2(x) \chi_{\mu\nu}(x) dx^{\mu} dx^{\nu}  \equiv a_U^2(x) \chi_{\mu\nu}^U (x) dx^{\mu} dx^{\nu}  \, ,\nonumber \\ 
& & \chi_{\mu\nu}^U (x) =\chi_{\mu}^{U\, a}(x) \chi_{\nu}^{U\, b}(x) \eta_{ab} \, , \quad \chi^U = \det \chi_{\mu}^{U\, a} = 1
\ee
which sets a typical basis for the Gravifield space-time. Such a basis is referred as a {\it Unitary basis}. The Unitary basis will be shown to be a physically meaningful basis for considering quantum effect within the framework of QFT. For convenience, the label $ ``U"$ for all the quantities in the Unitary basis will be omitted in the following discussions, the line element is expressed as 
\be
l_{\chi}^2 = a^2(x) \chi_{\mu\nu}(x) dx^{\mu} dx^{\nu} \, ; \qquad \chi = \det \chi_{\mu}^{\;\; a} = 1  \, .
\ee
Here the conformal scale field $a(x)$ (or conformal scalar field $\phi(x) \equiv M_S a(x)$) reflects directly the physics degree of freedom in the Unitary basis. Thus the scalar particle $\phi(x)$ in the Unitary basis may be called as {\it scalinon} particle that characterizes the conformal scaling evolution of universe.

\subsection{Conformal Proper Time and Cosmological Horizon}

The background structure of Gravifield Space-time is shown be governed by the background scalar field in the Unitary basis ($\chi = \det \chi_{\mu}^{\;\; a} = 1$)
\be 
& & \langle \chi_{\mu}^{\;\; a}(x) \rangle = \chib_{\mu}^{\;\; a}(x) = \eta_{\mu}^{\;\; a}  \, , \quad \langle \phi(x) \rangle = \bar{\varphi}(x)\, , \nonumber \\
& &  \langle \Om_{\mu}^{ab}(x) \rangle =  g_s^{-1} \partial^{\rho} \ln \pb(x)\, \eta_{\mu\rho}^{[ab]} \, , \quad  \langle W_{\mu} \rangle = g_w^{-1} \partial_{\mu}\ln \pb(x)\, .
\ee
which forms a {\it background Gravifield Space-time}  with the line element 
\be
\langle l_{\chi}^2 \rangle = \langle a^2(x) \eta_{ab}\, \chi^a(x) \chi^b(x) \rangle = \ab^2(x)\, \eta_{\mu\nu} \, dx^{\mu} dx^{\nu}  \, . 
\ee
Such a  background Gravifield Space-time coincides with a conformally flat Minkowski space-time governed by the background conformal scale field $\ab(x)$, which distinguishes from the globally flat Minkowski space-time that is introduced as an inertial reference frame of coordinates. $\ab(x)$ is  determined by the solutions of the background scalinon field $\pb(x)$ in the Unitary basis 
\be
\ab(x) \to \pm \ab_{\kappa\pm}(x) = \pm \pb_{\kappa\pm}(x)/M_S = \pm \frac{\mk }{\alpha_S M_S} \frac{1}{1 \mp x^{\mu}\kappa_{\mu}}\, .
\ee
which shows that the line element is invariant under the global Lorentz and scaling transformations. 

The scalar product  $x^{\mu}\kappa_{\mu}$ is a Lorentz and scaling invariant quantity,  which enables us to define a {\it conformal proper time} $\eta$ 
\be
x^{\mu}\kappa_{\mu} \equiv \kh \, c \eta  \, ,
\ee 
with $\kh$ as a {\it conformal proper energy scale} and $c$ the speed of light in vacuum. $\eta$ and $\kh$ obey a general conformal scaling transformation 
\be
\eta\to \eta' = \lambda^{-\alpha}\, \eta\, , \quad \kh \to \kh' = \lambda^{\alpha} \, \kh \, ,
\ee
with $\alpha$ an arbitrary constant parameter.  

By expressing the Lorentz time component $x^0=ct $ in terms of the {\it conformal proper time} $\eta$, we have
\be 
& & x^0 =ct = \frac{\kh}{\kappa_0}\, c \eta - \frac{\kappa_i}{\kappa_0} x^i \equiv  \frac{1}{u_0} ( c \eta - u_i x^i ) \, , \nonumber \\
& & u_0 \equiv \frac{\kappa_0}{\kh }\, , \quad  u_i \equiv  \frac{\kappa_i }{\kh}\, .
\ee  
The infinitesimal displacement $dx^0$ of the Lorentz time component is replaced by
\be 
dx^0 = c\, dt = \frac{1}{u_0} ( c d\eta - u_i dx^i ) \, ,
\ee 
and the conformal scale field is rewritten as
\be
\ab(x) \to \ab(\eta)\to \pm \ab_{\kappa\pm} (\eta) = \pm \frac{\mk}{\alpha_S M_S } \frac{1}{1 \mp \kh\,c \eta}\, .
\ee 
Thus the line element reads
\be
\langle l_{\chi}^2 \rangle 
& = & \ab^2(\eta) \chib_{\mu\nu} d\hat{x}^{\mu}d\hat{x}^{\nu}\, , \quad \hat{x}^{\mu} = (c\eta, x^i) \, , 
\ee
where $\chib_{\mu\nu}$ is a non-diagonal background metric for the background Gravifield space-time 
\be
\chib_{\mu\nu} = \frac{1}{u_0^2}
\left(
\begin{array}{cc}
 1 &  -u_j     \\
 -u_i  &    u_0^2\,  \chib_{ij}
\end{array}
\right)   \, , \quad 	 \chib_{ij} = \eta_{ij} + \frac{u_iu_j}{u_0^2 } \, .
\ee
 
The Lorentz and scaling invariance allows one to choose in principle a special reference frame and conformal scaling factor, so that $u_i =0$ ($\kappa_i =0$) and $\kh = \mk=\kappa_0$ and $u_0 =1$, which is known as the co-moving reference frame. Thus only in the co-moving reference frame, we arrive at an isotropic and homogenous background Gravifield space-time 
\be
\langle l_{\chi}^2 \rangle & = &   \ab^2(\eta)  \eta_{\mu\nu} d\hat{x}^{\mu}d\hat{x}^{\nu} =  \ab^2(t)  \eta_{\mu\nu} dx^{\mu}dx^{\nu}  \, .
\ee
In this case,  the conformal proper time $\eta$ coincides with the {\it co-moving Lorentz time} ($\eta = t$). 
 
It is interesting to note that the background conformal scale field has a singularity when the {\it conformal proper time} or the co-moving Lorentz time approaches to the epoch given by the inverse of {\it the conformal proper energy scale}, i.e., 
\be
\eta \to \ek \equiv \pm \frac{1}{c\kh}\, , \quad \ab(\eta)\to \ab_{\pm}(\ek) = \infty \, ,
\ee
at which the conformal size of  the background Gravifield space-time becomes infinitely large.  Thus the {\it cosmological horizon} in the background Gravifield space-time is defined by the light traveled distance 
\be
\Lk = c|\ek|  = 1/\kh \, .
\ee

In summary, the {\it background Gravifield space-time} is a conformally flat Minkowski space-time described by the {\it cosmic vector} $\kappa_{\mu}$ with the {\it cosmological mass} scale $\mk = \sqrt{\kappa_{\mu}\kappa^{\mu}}$. In terms of  {\it conformal proper time } $\eta = x_{\mu}\kappa^{\mu}/c\kh $, the background Gravifield space-time is characterized by the {\it cosmological horizon} $\Lk=1/\kh$ at which the {\it conformal scale factor} $\ab(\eta)$ becomes {\it singular}. 

\subsection{Cosmic Proper Time and Inflationary Universe}

In the conformally flat  background Gravifield space-time,  the conformal size of both the conformal proper time and space is expanded simultaneously. To describe the evolution of universe, it is useful to define a {\it cosmic proper time } $\tau$ which is related to the conformal proper time as follows
\be
d\tau \equiv \ab(\eta) d\eta  \, , \quad  \ab(\eta) \to \pm \ab_{\kappa\pm} (\eta) = \pm \frac{\mk}{\alpha_S M_S } \frac{1}{1 \mp \kh \, c\eta} = \pm \frac{\mk}{\alpha_S M_S } \frac{1}{1 \mp c\eta/\Lk }\, .
\ee 
After carrying out the integration, we obtain a relation between the {\it conformal proper time} $\eta$ and the {\it cosmic proper time} $\tau$
\be
 & & \pm \frac{1}{1 - \kh \, c\eta} = \pm e^{\Mk c \tau}\, , \quad -\infty \leq c\eta \leq \Lk \, , \quad  -\infty \leq \tau \leq \infty  \, , \nonumber \\
 & & \pm \frac{1}{1 - \kh \, c\eta} = \mp e^{-\Mk c \tau}\, , \quad \Lk  \leq c\eta \leq \infty \, , \quad  -\infty \leq \tau \leq \infty  \, , \nonumber \\
 & & \pm \frac{1}{1 + \kh \, c\eta} = \mp e^{\Mk c \tau}\, , \quad -\infty \leq c\eta \leq -\Lk \, , \quad  -\infty \leq \tau \leq \infty  \, , \nonumber \\
 & & \pm \frac{1}{1 + \kh \, c\eta} = \pm e^{-\Mk c \tau}\, , \quad -\Lk \leq c\eta \leq \infty \, , \quad  -\infty \leq \tau \leq \infty  \, . 
 \ee
which shows that it takes an infinitely large {\it cosmic proper time} $\tau \to \infty$ for the light traveling to the {\it cosmological horizon} $c\eta \to \Lk$. In obtaining the above relation, we have used the integration condition 
\[ \ab(\eta=0) = \ab(\tau=0) \] \, . 

The conformal scale factor for the spatial coordinates gets an exponential form in terms of the {\it cosmic proper time}
\be
& &  \ab(\tau) \to \pm \ab_{\kappa\pm}(\tau) = \pm \ab_0 e^{\pm \Mk c \tau} \, (\; \mbox{or} \; \mp \ab_0 e^{\mp\Mk c \tau} ) \, , \quad  \ab_0 = \kh/\Mk = \lk/\Lk \, ,
\ee
with the definitions
 \be 
 & & \Mk \equiv \alpha_S M_S \, \frac{\kh}{\mk} \, ,  \qquad \lk = \frac{1}{\Mk} \, .
 \ee
Here $\Mk$ and $\lk$ represent the {\it primary cosmic energy scale} and {\it primary cosmic horizon}, respectively. 

 The line element in terms of the {\it cosmic proper time} reads
\be
\langle l_{\chi}^2 \rangle & = & \gb_{\mu\nu}(\tau) d\hat{x}^{\mu}d\hat{x}^{\nu} \,  ; \qquad \hat{x}^{\mu} = (c\tau, x^i) \, ,
\ee
with the metric tensor 
\be
& & \bar{g}_{\mu\nu} (\tau) = \frac{1}{u_0^2}
\left(
\begin{array}{cc}
 1 &  - \ab(\tau)\, u_j     \\
 - \ab(\tau)\, u_i  &     \ab^2(\tau)\, u_0^2\, \bar{g}_{ij}
\end{array}
\right)   \, , \quad 	 \bar{g}_{ij} = \eta_{ij} + \frac{u_iu_j}{u_0^2 } \, ,
\ee
which indicates that the metric of the background Gravifield space-time is in general not isotropic in terms of the cosmic proper time $\tau$. Only in the co-moving reference frame with $u_i =0$ and $u_0=1$,  the {\it  cosmic proper time } $\tau$ is correlated to the {\it co-moving Lorentz time} $t$, the background Gravifield space-time appears to be isotropic and homogeneous. The line element is given by 
\be
\langle l_{\chi}^2 \rangle & = &  c^2 d\tau^2 +  \ab^2(\tau) \eta_{ij}\, dx^{i} dx^{j}  \, .
\ee
It is seen that the well-known form of Friedman-Lema"tre-Robertson-Walker (FLRW) metric is reproduced only in the specific co-moving reference frame. The resulting background Gravifield space-time characterizes an inflationary universe with $\ab(\tau) = \ab_0 e^{\Mk c \tau} $ (or deflationary universe with $\ab(\tau) = \ab_0 e^{-\Mk c \tau} $) when the cosmic proper time arrow goes from past to future.

\section{Quantization of Gravity Theory and Quantum Inflation}

Based on the background structure of Gravifield space-time after the gravitational gauge symmetry breaking, we shall discuss the quantization of gravitational gauge theory.  For the quantum effects, we will focus on the quantum inflation of early universe. 

\subsection{Quantization of Gravity }

As the global Lorentz and scaling invariant background Gravifield space-time is characterized by the conformally flat Minkowski space-time in the Unitary basis, all quantized fields are expressed as follows
\be
& & \chi_{\mu}^{\;\; a} = \chib_{\mu}^{\;\; a} + h_{\mu}^{\;\; a}(x)/M_W \, , \quad \chib_{\mu}^{\;\; a} = \eta_{\mu}^{\; a} \, , \nonumber \\
& &   \phi(x) =  \bar{\varphi}(x) + \varphi(x)\, , \quad \pb(x) = M_S\, \ab(x) \, , \nonumber \\
& &  \Om_{\mu}^{ab}(x) =  \tilde{\chi}_{\mu\rho}^{[ab]} \ob^{\rho}(x) +  \Omega_{\mu}^{ab}(x) = \eta_{\mu\rho}^{[ab]} \ob^{\rho}(x)  + \tilde{H}_{\mu\rho}^{[ab]} \ob^{\rho}(x)/M_W +  \Omega_{\mu}^{ab}(x) \, , \nonumber \\
& & W_{\mu}(x) = \bar{w}_{\mu}(x)  + w_{\mu}(x)\, , \quad g_w \wb_{\mu}(x) = g_s \ob_{\mu}(x) = \partial_{\mu} \ln \pb(x) \, , 
\ee
where we have introduced the following definitions
\be
 & & \tilde{\chi}_{\mu\rho}^{[ab]}(x) = \chi_{\mu}^{\;\; a} \hat{\chi}^b_{\;\;\rho}- \chi_{\mu}^{\;\; b}\hat{\chi}^a_{\;\;\rho}   = \langle \tilde{\chi}_{\mu\rho}^{[ab]}(x) \rangle + \tilde{H}_{\mu\rho}^{[ab]}(x) \, ,  \quad   \langle \tilde{\chi}_{\mu\rho}^{[ab]}(x) \rangle  = \eta_{\mu\rho}^{[ab]} \, , \nonumber \\
 & & \tilde{H}_{\mu\rho}^{[ab]}(x) = h_{\mu}^{\;\; a} \eta^b_{\;\;\rho} - h_{\mu}^{\;\; b} \eta^a_{\;\;\rho}
 - \eta_{\mu}^{\;\; a}\hat{h}^b_{\;\;\rho}  + \eta_{\mu}^{\;\; b}\hat{h}^a_{\;\;\rho}  - (\, h_{\mu}^{\;\; a} \hat{h}^b_{\;\;\rho} - h_{\mu}^{\;\; b} \hat{h}^a_{\;\;\rho} \, )/M_W \, , \nonumber \\
 & & \hat{\chi}_a^{\;\; \mu}(x) \equiv \eta_a^{\; \mu} - \hat{h}_a^{\; \mu}(x)/M_W \, , \quad \hat{h}_{a}^{\;\; \mu} \equiv h_a^{\;\; \mu} -  \Nh_a^{\;\; \mu}/M_W = \sum_{n=1}^{\infty}  \frac{(-1)^{n-1}}{M_W^{n-1}} (h^n)_{a}^{\;\; \mu}  \, .
 \ee
 with the condition in the Unitary basis ($\chi = \det \chi_{\mu}^{\;\; a}=1$) 
 \be 
& &  \ln \det \chi_{\mu}^{\;\; a} =   \Tr\ln ( \, \eta_{\mu}^{\; a} + h_{\mu}^{\;\; a}/M_W\, )  =0 \, , \;\; \mbox{or} \;\; \sum_{n=1}^{\infty} \frac{ (-1)^{n-1}}{ n M_W^{n-1}} (h^n)_{\mu}^{\;\; a} \eta_a^{\; \mu} =0 \, .
\ee
Here $M_W$ is introduced as a {\it weighting energy scale}  to make the quantized Gravifield $h_{\mu}^{\;\; a}(x)$ dimensional and fixed via the normalization of kinetic term for the Gravifield. With the convention for the conformal scale field $\phi(x) = M_S\, a(x)$,  $M_W$ can be expressed by the basic scaling energy scale $M_W^2 = \alpha_W M_S^2$.

The action in terms of the above quantized fields $h_{\mu}^{\;\; a}(x)$, $\varphi(x)$, $w_{\mu}(x)$, $\Omega_{\mu}^{ab}(x)$ gets the following form in the Unitary basis
\be
\label{action4}
S_{\chi}  & = & \int d^{4}x\;  \frac{1}{2}  [\,  \hat{\chi}^{\mu\nu} (\bar{\Psi} \chi_{\mu}^{\;\; a}\gamma_a i {\mathsf D }_{\nu}   \Psi  + \bar{\psi} \chi_{\mu}^{\;\; a}\gamma_a i \td_{\nu}   \psi )  + H.c. \, ]  - y_s \bar{\psi} (\pb + \varphi) \psi \nonumber \\
& - & \frac{1}{4}  \hat{\chi}^{\mu\mu'} \hat{\chi}^{\nu\nu'} [\, {\cal F}^I_{\mu\nu} {\cal F}^{I}_{\mu'\nu'}  + {\mathsf R}_{\mu\nu}^{ab}  {\mathsf R}_{\mu'\nu' ab}  + W_{\mu\nu} W_{\mu'\nu'}   \, ]   \nonumber \\
& + &  \alpha_W (\, \pb + \varphi \,)^2 \frac{1}{4}  \hat{\chi}^{\mu\mu'} \hat{\chi}^{\nu\nu'}  {\mathsf G}_{\mu\nu}^a {\mathsf G}_{\mu'\nu' a}  + \frac{1}{2} \hat{\chi}^{\mu\nu} d_{\mu} \varphi d_{\nu}\varphi  -  \lambda_s ( \pb + \varphi  )^4    \nonumber \\
& - & [\, \alpha_E ( \pb + \varphi )^2 - \hat{\chi}^{\rho\sigma} \ob_{\rho}\ob_{\sigma} \, ]  g_s \hat{\chi}_a^{\;\;\mu} \hat{\chi}_b^{\;\;\nu}  {\mathsf R}_{\mu\nu}^{ab} \nonumber \\
& + & 6 \alpha_E g_s ( \pb + \varphi  )^2  [\,  \hat{\chi}_a^{\;\mu}  \td_{\mu} (\hat{\chi}^{a\nu} \ob_{\nu} ) + g_s \hat{\chi}^{\mu\nu} \ob_{\mu}\ob_{\nu}  \, ]  \nonumber \\
&- & 2 \alpha_E g_s ( \pb + \varphi )^2 \hat{\chi}^{\nu\rho} \ob_{\rho} \hat{\chi}_a^{\;\;\mu}  G_{\mu\nu}^a  - 3 g_s^2  (\, \hat{\chi}^{\mu\nu} \ob_{\mu}\ob_{\nu} )^2  \nonumber \\
& + &  \frac{1}{2} \hat{\chi}^{\mu\nu}[\, g_w^2 ( w_{\mu}\pb  + \wb_{\mu} \varphi) ( w_{\nu}\pb  + \wb_{\nu} \varphi) - 2g_w (\pb w_{\mu} + \wb_{\mu} \varphi) d_{\nu} \varphi \, ] \nonumber \\
& - & \hat{\chi}_b^{\;\rho}\, \ob_{\rho} \,  \hat{\chi}^{\mu\mu'} \hat{\chi}^{\nu\nu'} G_{\mu'\nu' a}  {\mathsf R}_{\mu\nu}^{ab}  -2\hat{\chi}_a^{\;\;\mu} \hat{\chi}^{\nu\nu'}   [ \, \td_{\nu'}(\hat{\chi}_b^{\;\; \rho}\ob_{\rho})  - 
g_s \ob_{\nu'}  \hat{\chi}_b^{\;\; \rho}\ob_{\rho}  \, ]  {\mathsf R}_{\mu\nu}^{ab}    \nonumber \\
& - & \frac{1}{4}  \hat{\chi}^{\mu\mu'} \hat{\chi}^{\nu\nu'}[\,  \hat{\chi}^{\rho\sigma}\ob_{\rho}\ob_{\sigma}\eta_{ab} -  \hat{\chi}_a^{\;\;\rho} \hat{\chi}_b^{\;\;\sigma} \ob_{\rho}\ob_{\sigma}\, ] G_{\mu\nu }^a G_{\mu'\nu'}^b \nonumber \\ 
& - & 2 \hat{\chi}^{\mu\nu} \td_{\mu} ( \hat{\chi}_a^{\;\;\rho} \ob_{\rho}) \td_{\nu} ( \hat{\chi}^{a \sigma} \ob_{\sigma}) - 
[\hat{\chi}_a^{\;\mu}  \td_{\mu} ( \hat{\chi}^{a \rho} \ob_{\rho}) ]^2  \nonumber \\
& - & \hat{\chi}^{\mu\mu'} \hat{\chi}^{\nu\nu'}  G_{\mu\nu }^a  \ob_{\mu'} \td_{\nu'}(\hat{\chi}_a^{\;\; \rho}\ob_{\rho}) 
 + \frac{1}{2} \hat{\chi}_a^{\;\;\mu}  \hat{\chi}^{\nu\nu'}  G_{\mu\nu }^a  \partial_{\nu'}(\hat{\chi}^{\rho\sigma}\ob_{\rho} \ob_{\sigma})  \nonumber \\
 & + & 2 g_s \hat{\chi}^{\mu\nu}  \partial_{\mu}(\hat{\chi}^{\rho\sigma}\ob_{\rho} \ob_{\sigma}) \ob_{\nu} 
 - 4g_s \hat{\chi}_a^{\;\mu}  \td_{\mu}(\hat{\chi}^{a \nu}\ob_{\nu})  \hat{\chi}^{\rho\sigma}\ob_{\rho} \ob_{\sigma}    +  {\cal L}'(x)   \, , 
\ee
with 
\be
& & i{\mathsf D}_{\mu} = i\partial_{\mu} + \Omega_{\mu} + {\cal A}_{\mu}  \nonumber \\
& & {\mathsf R}_{\mu\nu}^{ab}  = \partial_{\mu}\Omega_{\nu}^{ab} -   \partial_{\nu}\Omega_{\mu}^{ab} + g_s( \Omega_{\mu c}^{a} \Omega_{\nu}^{cb}  - \Omega_{\nu c}^{a} \Omega_{\mu}^{cb} )  \nonumber \\
& & W_{\mu\nu} = \partial_{\mu} w_{\nu} - \partial_{\nu} w_{\mu}  \, , \quad d_{\mu}\varphi = (\partial_{\mu} - g_w w_{\mu} ) \varphi \nonumber \\
& & {\mathsf G}_{\mu\nu}^a = \nabla_{\mu} \chi_{\nu}^{\;\; a}  - \nabla_{\nu} \chi_{\mu}^{\;\; a}   =  G_{\mu\nu}^a +   g_w (w_{\mu} \chi_{\nu}^{\;\; a} -  w_{\nu} \chi_{\mu}^{\;\; a} )   \, ,\nonumber \\
& & \hat{\chi}^{\mu\mu'} \hat{\chi}^{\nu\nu'}  {\mathsf G}_{\mu\nu}^a {\mathsf G}_{\mu'\nu' a} = \hat{\chi}^{\mu\mu'} \hat{\chi}^{\nu\nu'}  G_{\mu\nu}^a G_{\mu'\nu' a} + 4g_w \hat{\chi}^{\mu\rho}  \hat{\chi}_a^{\;\;\nu} G_{\mu\nu}^a w_{\rho} 
+ 6 g_w^2 \hat{\chi}^{\mu\nu} w_{\mu}w_{\nu} \, ,  \nonumber \\
& & G_{\mu\nu}^a = \td_{\mu} \chi_{\nu}^{\;\; a}  - \td_{\nu} \chi_{\mu}^{\;\; a}   = \chi_{\mu\nu}^a +    g_s (\Omega_{\mu b}^a \chi_{\nu}^{\;\; b}  -  \Omega_{\nu b}^a \chi_{\mu}^{\;\; b} )\, . 
\ee
It is shown that there exist rich gravitational interactions with respect to the background fields. Note that  all the linear terms of quantized fields become vanishing and the gravitational gauge symmetries are broken down to the corresponding global symmetries when substituting the background field solutions obtained from the equations of motion.

\subsection{Massless Graviton and Massive Spinnon}

Practically, it is convenient to work in Euclidean space-time by making a Wick rotation.  Classically, the Gravifield $\chi_{\mu}^{\;\; a}(x)$ has sixteen field components, $\Omega_{\mu}^{ab}(x)$ twenty four field components and $W_{\mu}(x)$ four field components.  There are totally forty five components including single scalar field. For  the massless gauge fields, it is known that only the transverse components are physical ones. There are eight, twelve and two independent physical degrees of freedom for the massless gauge-type Gravifield $\chi_{\mu}^{\;\; a}(x)$,  spinnic gauge field $\Omega_{\mu}^{ab}(x)$ and scaling gauge field $W_{\mu}(x)$, respectively. When fixing the scaling gauge condition to the unitary basis, i.e., $\chi = \det \chi_{\mu}^{\;\; a}(x) = 1$,  the Gravifield $\chi_{\mu}^{\;\; a}(x)$ is reduced to have seven independent physical degrees of freedom.  Thus there are twenty two independent physical degrees of freedom including the single scalar field.

After considering the gravitational gauge symmetry breaking with $\langle \chi_{\mu}^{\;\; a}(x) \rangle = \chib_{\mu}^{\;\; a} = \eta_{\mu}^{\;\; a} $, $\langle \phi(x) \rangle = \pb(x) = \ab M_S $, the leading Gravifield dynamical interaction term provides masses for the spinnic and scaling gauge fields   
\be 
{\cal L}_G^{(m)} = \frac{3}{2}\ab^2 g_w^2 M_W^2 w_{\mu} w_{\nu} \eta^{\mu\nu}  + \frac{1}{4} \ab^2 g_s^2 M_W^2 
  \Omega_{[\mu a b] } \Omega_{[\mu' a' b']} \eta^{\mu\mu'} \eta^{aa'}   \eta^{bb'}  \, ,
\ee
with $\Omega_{[\mu a b]} \equiv  \Omega_{\mu a b } - \Omega_{b a \mu }$.

It shows that the totally antisymmetric part of the spinnic gauge field  $\Omega_{[\mu a b]} $ become massive. The massive spinnic gauge field yields additional four physical degrees of freedom and has totally sixteen physical degrees of freedom. The massive scaling gauge field possesses three physical degrees of freedom. Here the gauge-type Gravifield plays the role as a Higgs-type boson for generating the spinnic and scaling gauge symmetry breaking and leading to the massive spinnic and scaling gauge fields. Thus its five physical degrees of freedom will be eaten by the spinnic and scaling gauge fields. Eventually,  the gauge-type Gravifield has only two independent physical degrees of freedom, which is naturally chosen to be the following two transverse components of Gravifield $\chi_{\mu}^{\;\; a}(x) = \eta_{\mu}^a + h_{\mu}^{\;\; a}(x) /M_W$ 
\be \label{Graviton1}
h_{12}(x) = h_{21}(x), \quad h_{11}(x) = - h_{22}(x) \, .
\ee
Such a massless quantized transverse Gravifield $h_{ij}$ acts as  the {\it graviton}. The totally antisymmetric spinnic gauge field $\Omega_{[\mu a b]}$ acts as a massive {\it spinnon}, it provides a twisting and torsional interaction.

\subsection{Quantum Inflation of Early Universe}

The solution for the conformal scale field in the background Gravifield space-time shows that the evolution of early universe should undergo an extremely rapid exponential expansion in terms of the cosmic proper time.  The main issue is what causes the inflation of the universe and which epoch the inflationary universe gets end. We shall show that the quantum effect plays a significant role to answer the question.
 
The Lagrangian density and energy-momentum tensor for the background Gravifield space-time after gravitational gauge symmetry breaking are found to be 
\be
\bar{{\cal L}}  & = &  - 6 \partial_{\mu} \ob_{\nu} \partial^{\mu} \ob^{\nu} + 12 \alpha_E g_s^2 \ob_{\mu}^2 \pb^2 - \lambda_s \pb^4 =  - 6 g_s^2 \ob_{\mu}^2 \ob_{\mu}^2 + 12 \alpha_E g_s^2 \ob_{\mu}^2 \pb^2 - \lambda_s \pb^4  \, , \nonumber \\
& = &  - ( 6\alpha_S^4/g_s^2 - \lambda_s  )\, \pb^4 =  - \lambda_s (\,\alpha_S^2\alpha_E^{-1}/ g_s^2 -1  \, )\, \pb^4  \, , \\
\bar{T}_{\mu\nu} &  = & - \eta_{\mu\nu}2[\, d_{\rho}\ob_{\sigma} d^{\rho}\ob^{\sigma} + 2 g_s d_{\rho}\ob^{\rho} \ob_{\sigma}\ob^{\sigma} 
+ 3 g_s^2  (\ob_{\sigma}\ob^{\sigma})^2 - \alpha_W \Ob_{\sigma}\Ob^{\sigma}\pb^2 \, ] \nonumber \\
& & \;\;\; - \eta_{\mu\nu} 2 \alpha_Eg_s (d_{\rho}\ob^{\rho} + 3g_s \ob_{\rho}\ob^{\rho})  \pb^2 
- [\,  \bar{W}_{\mu\sigma} \bar{W}_{\nu}^{\sigma} -  d_{\mu}\pb d_{\nu}\pb  - 2 \alpha_W  \Ob_{\mu} \Ob_{\nu} \pb^2 ] \nonumber \\
& &\;\;\; - 2[\, d_{\mu}\ob_{\sigma}  d_{\nu}\ob^{\sigma}  +  (d_{\mu}\ob_{\nu}  + d_{\nu}\ob_{\mu} ) d_{\sigma}\ob^{\sigma} - d_{\sigma}\ob_{\mu} d^{\sigma}\ob_{\nu} + 2 g_s  (d_{\mu}\ob_{\nu} + d_{\nu}\ob_{\mu}  ) \ob_{\sigma}\ob^{\sigma} \, ] \nonumber \\
& & \;\;\; +  4 \alpha_Eg_s \pb^2 d_{\nu}\ob_{\mu} -\eta_{\mu\nu}\bar{\cal L} = 0 \, , \label{EMT2}
\ee 
where the final expression for $\bar{{\cal L}}$ and the result $\bar{T}_{\mu\nu} =0$ are resulted by applying for the solutions of the background fields.  It indicates that the background Gravifield space-time represents the whole universe as it has no energy-momentum exchanging with its exterior.    

The stability of the background Gravifield space-time requires a minimal background scalar potential with 
\be 
& & \pb \to 0\, , \; \; \qquad \mbox{for} \quad   6\alpha_S^4/g_s^2  >  \lambda_s = 6 \alpha_E\alpha_S^2  > 6 \alpha_E^2 g_s^2  \, , \nonumber \\
& & \pb \to \pm \infty \, , \; \quad \mbox{for} \quad 6\alpha_S^4/g_s^2 <   \lambda_s = 6 \alpha_E\alpha_S^2  < 6 \alpha_E^2 g_s^2  \, .
\ee
It is seen that the early universe characterized by the background Gravifield space-time without considering quantum effect will be stabilized either at an infinitely small conformal size $\pb \to 0 $ for $ \alpha_S^2 > \alpha_E g_s^2 $ ($ \alpha_S^4  >  \lambda_s g_s^2/6 $) or at an infinitely large conformal size $\pb \to \pm \infty$ for $ \alpha_S^2  < \alpha_E g_s^2 $ ($ \alpha_S^4  < \lambda_s g_s^2/6 $). 

Let us consider that our universe has a beginning. The background Gravifield space-time must be stabilized at an infinitely small conformal size with $\ab \ll 1$. For an extreme small size of universe at the beginning, the quantum effect must be considered. For our present purpose, we shall focus on the possible one loop quantum effects on the background Gravifield space-time. The leading contributions to the effective Lagrangian of the scalinon field gets the following general form 
  \be
  {\cal L}_S & \simeq & \mu_U^2\, (\, \pb(x) + \varphi(x)\, )^2 - \lambda_U\, (\, \pb(x) +\varphi(x)\, )^4  \nonumber \\ 
  & + &  2\lambda_s (\, \pb(x) + \varphi(x)\, )^2 \pb^2(x)
  +  \bar{\mu}_U^2\, \pb^2(x) - \bar{\lambda}_U\, \pb^4(x) \,  , 
  \ee
with 
\be
& & \mu_U^2 (M_U/\mu) \simeq \frac{2}{(4\pi)^2} y_S^2 \left(M_U^2-\mu^2 \right) \,  , \nonumber \\
& & \lambda_U(M_U/\mu)  \simeq  \lambda_s + \frac{4}{(4\pi)^2} \lambda_S^2 \, \ln \frac{M_U^2}{\mu^2}  \, , \nonumber \\
& & \bar{\mu}_U^2 (M_U/\mu) \simeq \frac{2}{(4\pi)^2} \bar{y}_S^2 \left(M_U^2-\mu^2 \right)\, , \nonumber \\
& & \bar{\lambda}_U(M_U/\mu)  \simeq   \frac{\alpha_S^2}{g_s^2\alpha_E}  \lambda_s  + \frac{4}{(4\pi)^2} \bar{\lambda}_S^2 \, \ln \frac{M_U^2}{\mu^2}  \, , 
\ee
with
\be
& &  y_S^2 \equiv y^2_s- 6\lambda_s- c_w g_w^2 -c_s g_s^2  \, , \nonumber \\
 & & \lambda_S^2  \equiv  -12 \lambda_s^2 - \left( \lambda_s - 2 y_s^2\right) y_s^2+ c'_w g_w^4 + c'_s g_s^4  \, , \nonumber \\
 & & \bar{y}_S^2 \equiv \lambda_s +  \bar{c}_s \alpha_S^2  \, , \qquad \bar{\lambda}_S^2 \equiv -\lambda_s^2 +  \bar{c}'_s \alpha_S^4  \, ,
\ee
$M_U$ is taken to be a basic energy scale of QFT in the ultraviolet (UV) region and $\mu$ is the sliding energy scale set to be the physically interesting energy scale. The effective parameters $\mu_U^2$, $\lambda_U$, $\bar{\mu}_U^2$ and $\bar{\lambda}_U$ receive contributions from loops of the singlet fermion field and scalinon field characterized by the terms proportional to the parameters $y_s$ and $\lambda_s$, respectively.  The contributions from the scaling and spinnic gauge fields are given by the terms proportional to the parameters $g_w$ and $g_s$ or $\alpha_S$. $c_w$, $c'_w$, $c_s$, $c'_s$, $\bar{c}_s$ and $\bar{c}'_s$ represent the magnitudes of the contributions. Their values do not affect our present general considerations as $g_w$, $g_s$ and $\alpha_S$ are free parameters. 

The purely quantum induced effective mass parameters $\mu^2_U$ and $\bar{\mu}^2_U$ are attributed to the loop quadratic contributions. The magnitude of $\mu^2_U$ and $\bar{\mu}^2_U$ are characterized by the ultraviolet (UV) basic energy scale $M_U$ and  the sliding energy scale $\mu$ as well as the coupling constants at a given energy scale $\mu$. The quantum loop quadratic contributions have been shown to play a crucial role for understanding the electroweak symmetry breaking and hierarchy problem within the standard model of particle physics\cite{QEWSB}. Similarly, with the quantum induced positive mass parameter $\mu_U^2, \bar{\mu}_U^2  > 0$ with $y_S^2, \bar{y}_S^2 >0$, we yield an unstable potential which generates the inflation of early universe.  For a positive effective coupling parameter $\lambda_U , \bar{\lambda}_U > 0$, the effective potential for the scalinon field is spontaneously broken down to an evolving minimal vacuum, 
\be
  V_S^2(M_U/\mu) & = & \langle (\pb + \varphi)^2 \rangle   
  =  \frac{\mu_U^2 + \lambda_s \bar{\mu}_U^2/\bar{\lambda}_U  }{ 2( \lambda_U- \lambda_s^2/\bar{\lambda}_U  ) } \nonumber \\
  \bar{V}_S^2(M_U/\mu) & = & \langle \pb^2 \rangle   
  =  \frac{\bar{\mu}_U^2}{2\bar{\lambda}_U} +  \frac{\lambda_s}{\bar{\lambda}_U}\,  V_S^2(M_U/\mu)  \, , 
  \ee
which leads the inflationary universe to get end. One can always yield a solution to satisfy such requirements as the couplings $y_s$, $\lambda_s$, $g_s$, $g_w$ and $\alpha_S$ are all free parameters. 

It is noticed that that the loop quadratic contribution causes the breaking of the global scaling symmetry.The scalinon field not only behaves as a quantum field but also characterizes the conformal size of universe. At the beginning of universe, all quantum fields must have a high energy momentum at the UV basic energy scale $M_U$, while the smallness of the scalinon field characterized by the cosmological mass scale $m_{\kappa}$ indicates that a small variation of quantum fluctuation can cause a significant change for the scalinon field. Suppose that the variation of quantum fluctuation is at the order of the cosmological mass scale $m_{\kappa}$, namely the sliding energy scale is away from the UV energy scale by a very small amount with $\mu_{\kappa} \sim M_U - m_{\kappa}$, thus the eVEV will be stabilized around
\be
V_S & \equiv&  V_S(M_U/\mu_{\kappa} ) = \frac{1}{(4\pi)} \sqrt{ \frac{y_S^2 + \bar{y}_S^2 \lambda_s/\bar{\lambda}_U }{\lambda_U - \lambda_s^2/\bar{\lambda}_U  }   }\,  \sqrt{2M_Um_{\kappa} + m_{\kappa}^2 } \, , \nonumber \\
\bar{V}_S & \equiv &  \bar{V}_S(M_U/\mu_{\kappa} ) = \frac{1}{(4\pi)}   [\, \frac{\bar{y}_S^2}{ \bar{\lambda}_U  }    +  (\frac{\lambda_s}{\bar{\lambda}_U})\, \frac{y_S^2 + \bar{y}_S^2 \lambda_s/\bar{\lambda}_U }{\lambda_U - \lambda_s^2/\bar{\lambda}_U  }  \, ]^{1/2} \, \sqrt{2M_Um_{\kappa} + m_{\kappa}^2 } 
\ee
which determines the epoch when the inflationary universe gets end with the conformal scale factor
\be
\ab_e = \langle \ab(x) \rangle = \frac{\bar{V}_S}{M_S} =  \frac{\mk }{\alpha_S M_S} \frac{1}{1 - \kh\, c \eta_e } = \ab_0 e^{ \Mk c \tau_e}\, .
\ee
 
In conclusion, it is the quantum effect of loop quadratic contribution that generates the inflation of universe. A small variation of quantum fluctuation at the order of cosmological mass scale leads the inflationary universe to end as the quantum effect causes the breaking of global scaling symmetry and yields a nonzero eVEV of the background scalar potential. We may refer such an inflation of the universe as the {\it Quantum scalinon Inflation}.

\section{Space-time Gauge Field with Goldstone-like Gravifield $\&$ Gravimetric Field and Quantum Dynamics of Space-time }

The quantum dynamics of Gravifield space-time is characterized by the quantized Gravifield and conformal scalar field as well as spinnic and scaling gauge fields. To correlate the quantum dynamics property between the Gravifield space-time and  coordinate space-time, it is useful to settle a special gauge condition via the Gravifield in such a way that the spinnic gauge symmetry is transmuted into a hidden gauge symmetry.

\subsection{Space-time Gauge Field with Goldstone-like Gravifield $\&$ Gravimetric Field }

The quantum fields in the Unitary basis can be expressed in the Einstein-type basis by making a scaling gauge transformation 
\[ 
\chi_{\mu}^{\; a} (x) \to a_E(x) \chi_{\mu}^{\; a}(x)\, , \quad  \phi(x) \to  \phi(x)/a_E(x)\, , \quad g_w W_{\mu}(x) \to g_w W_{\mu}(x)  -  \partial_{\mu} \ln a_E(x) \, . \]
 As a consequence, the quantum fields get the following forms in the Einstein-type basis
\be
& & \chi_{\mu}^{\;\; a} (x) = a_E (x) [\, \eta_{\mu}^{\; a} + h_{\mu}^{\;\; a}(x)/M_W\, ] = ( \ab(x) + \varphi(x)/M_S)  [\, \eta_{\mu}^{\; a} + h_{\mu}^{\;\; a}(x)/M_W\, ]  \, , \nonumber \\
& &  \hat{\chi}_a^{\;\; \mu}(x)  \equiv a_E^{-1}(x) [\, \eta_a^{\; \mu} - \hat{h}_a^{\; \mu}(x)/M_W \, ] = ( \ab(x) + \varphi(x)/M_S)^{-1}  [\, \eta_{\mu}^{\; a} + h_{\mu}^{\;\; a}(x)/M_W\, ]^{-1}  \, , \nonumber \\
& & \chi (x) = \det \chi_{\mu}^{\;\; a}(x) = a_E^4(x) = ( \ab(x) + \varphi(x)/M_S)^4 \, , \quad   \phi(x) =   M_S \, , \nonumber \\
& &  \Om_{\mu}^{ab}(x) =  \tilde{\chi}_{\mu\rho}^{[ab]} \ob^{\rho}(x) +  \Omega_{\mu}^{ab}(x) \, , \quad g_s \ob_{\mu}(x) = \partial_{\mu} \ln \ab(x) \, , \nonumber \\
& & W_{\mu}(x) = \bar{w}_{\mu}(x)  + w_{\mu}(x)\, , \quad  \wb_{\mu}(x) = 0\, .
\ee
Here we shall omit the label ``E'' in all quantities for convenience.

Let us now construct a space-time gauge field through the Gravifield and spinnic gauge field as follows
\be
\Gm_{\mu\nu}^{\sigma}(x)  =  \hat{\chi}_a^{\; \sigma} \td_{\mu} \chi_{\nu}^{\; a} = \hat{\chi}_a^{\; \sigma} (\partial_{\mu} \chi_{\nu}^{\; a} + g_s \Omega_{\mu b}^{a} \chi_{\nu}^{\; b}  ) \, .
\ee
Here the Gravifield appears as a Goldstone-like field that transmutes the local spinnic gauge symmetry into the global Lorentz symmetry, so that the spinnic gauge symmetry becomes a hidden symmetry. The gauge field $\Gm_{\mu\nu}^{\sigma}(x) $ is a Lorentz tensor field defined in the flat Minkowski space-time. We now decompose the spinnic gauge field into two parts with the following properties under the spinnic gauge transformation $S(x)= e^{i\alpha_{ab}(x) \Sigma^{ab} /2} \in SP(1,3)$ 
\be
& & \Omega_{\mu}^{ab}(x)  =   \omega_{\mu}^{ab}(x) +  \ot_{\mu}^{ab}(x) \, \nonumber \\
 & & \omega_{\mu}(x) \to \omega'_{\mu}(x) = S \omega_{\mu}(x) S^{-1} + S \partial_{\mu} S^{-1} \, , \quad \omega_{\mu}(x) = \omega_{\mu}^{ab}(x) \frac{1}{2} \Sigma_{ab} \, , \nonumber \\
& & \ot_{\mu}(x) \to \ot'_{\mu}(x) = S \ot_{\mu}(x) S^{-1}\, , \quad \ot_{\mu}(x) = g_s \ot_{\mu}^{ab}(x) \frac{1}{2} \Sigma_{ab}\, .
 \ee
 which divides the {\it space-time gauge field } $\Gm_{\mu\nu}^{\sigma}(x)$ into two gauge invariant parts 
 \be
& & \Gm_{\mu\nu}^{\sigma}(x) \equiv  \Gamma_{\mu\nu}^{\sigma}(x) + \Gmt_{\mu\nu}^{\sigma}(x) \, , \nonumber \\
& &  \Gamma_{\mu\nu}^{\sigma} (x)  \equiv \hat{\chi}_a^{\; \sigma} (\partial_{\mu} \chi_{\nu}^{\; a} + g_s \omega_{\mu b}^{a} \chi_{\nu}^{\; b}  ) \, , \nonumber \\
& & \Gmt_{\mu\nu}^{\sigma}(x) \equiv g_s \hat{\chi}_a^{\; \sigma} \ot_{\mu b}^{a} \chi_{\nu}^{\; b} 
= - g_s \frac{1}{2}  \ot_{\mu ab} \tilde{\chi}_{\nu\rho}^{[a b]} \eta^{\rho\sigma} \, .
\ee
From the above definition, it can be shown that $\Gmt_{\mu\nu}^{\sigma}(x)$ constitutes an antisymmetric part of the {\it space-time gauge field }  $\Gm_{\mu\nu}^{\sigma}(x)$ with the following relation
\be
\Gmt_{\mu \nu}^{\sigma'}(x)\, \chi_{\sigma' \sigma} = - \Gmt_{\mu \sigma}^{ \nu'}(x)\, \chi_{\nu' \nu} \, .
\ee
By requiring $\Gamma_{\mu\nu}^{\sigma}(x)$ to be a symmetric part $ \Gamma_{\mu\nu}^{\sigma}(x) =  \Gamma_{\nu\mu}^{\sigma}(x)  $, we are able to show that the spinnic gauge field $\omega_{\mu}^{ab}(x) $ is characterized purely by the {\it Goldstone-like Gravifield} $ \chi_{\mu}^{\;a} $ with the following explicit form
\be
  g_s \omega_{\mu}^{ab}(x) & = &  \hat{\chi}^{a\nu} \chi_{\mu\nu}^b - \hat{\chi}^{b\nu} \chi_{\mu\nu}^a -  \hat{\chi}^{a\rho}  \hat{\chi}^{b\sigma}  \chi_{\rho\sigma}^c \chi_{\mu c }\, , \quad   \chi_{\mu\nu}^a = \partial_{\mu} \chi_{\nu}^{\; a} -  \partial_{\nu} \chi_{\mu}^{\;a} \, .
\ee
Taking such a pure gauge-like field for $\omega_{\mu}^{ab}(x) $,   $\Gamma_{\mu\nu}^{\sigma}(x) $ is found to be 
\be
& &  \Gamma_{\mu\nu}^{\sigma} = \Gamma_{\nu\mu}^{\sigma} = \frac{1}{2} \hat{\chi}^{\sigma\lambda} [\, \partial_{\mu} \chi_{\nu \lambda} + \partial_{\nu} \chi_{\mu \lambda} - \partial_{\lambda}\chi_{\mu\nu} \, ]   \, , \nonumber \\
& &  \chi_{\mu\nu}(x) = \chi_{\mu}^{\; a}(x)\chi_{\nu}^{\; b}(x)\eta_{ab}\, , \qquad \hat{\chi}^{\mu\nu}(x) = \hat{\chi}_a^{\;\mu}(x) \hat{\chi}_b^{\; \nu}(x)\eta^{ab}  \, .
\ee
$\Gamma_{\mu\nu}^{\sigma}(x) $ is determined completely by the tensor field $\chi_{\mu\nu}(x)$ (or its inverse $\hat{\chi}^{\mu\nu}(x)$) to be a pure gauge-like {\it space-time gauge field}. Where $\chi_{\mu\nu}(x)$ is regarded as a {\it Goldstone-like Gravimetric field} of space-time with ten degrees of freedom. 

Correspondingly,  the field strength $\Rm_{\mu\nu\rho}^{\sigma} $ can be decomposed into two parts
\be
& & \Rm_{\mu\nu\rho}^{\sigma} = R_{\mu\nu\rho}^{\;\sigma}  + \Rmt_{\mu\nu\rho}^{\sigma} \, , \nonumber \\
& & R_{\mu\nu\rho}^{\;\sigma}  = \partial_{\mu} \Gamma_{\nu\rho}^{\sigma} - \partial_{\nu} \Gamma_{\mu\rho}^{\sigma} -
\Gamma_{\mu\rho}^{\lambda} \Gamma_{\nu\lambda}^{\sigma} + \Gamma_{\nu\rho}^{\lambda} \Gamma_{\mu\lambda}^{\sigma}\, ,\nonumber \\
& & \Rmt_{\mu\nu\rho}^{\;\sigma}  = \nabla_{\mu} \Gmt_{\nu\rho}^{\sigma} - \nabla_{\nu} \Gmt_{\mu\rho}^{\sigma} -
\Gmt_{\mu\rho}^{\lambda} \Gmt_{\nu\lambda}^{\sigma} + \Gmt_{\nu\rho}^{\lambda} \Gmt_{\mu\lambda}^{\sigma} \, , 
\ee
with
\be
& & \nabla_{\mu} \Gmt_{\nu\rho}^{\sigma} = \partial_{\mu} \Gmt_{\nu\rho}^{\sigma} - \Gamma_{\mu\rho}^{\lambda}  \Gmt_{\nu\lambda}^{\sigma} + \Gamma_{\mu\lambda}^{\sigma} \Gmt_{\nu\rho}^{\lambda} \, .
\ee
The field strength of the space-time gauge field is related to the field strength of the spinnic gauge field as follows
\be
g_s{\mathsf R}_{\mu\nu}^{ab}  & = & \frac{1}{2} \Rm_{\mu\nu\rho}^{\sigma}\, \tilde{\chi}_{ \sigma\rho'}^{[ab]} \eta^{\rho'\rho}
=  \frac{1}{2} (R_{\mu\nu\rho}^{\sigma} +  \Rmt_{\mu\nu\rho}^{\sigma}  )\,  \tilde{\chi}_{ \sigma\rho'}^{[ab]} \eta^{\rho'\rho} \, .
\ee 
The field strength of Gravifield is found, in terms of the space-time gauge field, to be 
\be
G_{\mu\nu}^a & = & \btd_{\mu} \chi_{\nu}^{\;\; a}  - \btd_{\nu} \chi_{\mu}^{\;\; a}   =   \partial_{\mu} \chi_{\nu}^{\;\; a}  - \partial_{\nu} \chi_{\mu}^{\;\; a}  +    g_s (\Omega_{\mu b}^a \chi_{\nu}^{\;\; b}  -  \Omega_{\nu b}^a \chi_{\mu}^{\;\; b} ) \nonumber \\ 
& = &  ( \Gm_{\mu\nu}^{\sigma} - \Gm_{\nu\mu}^{\sigma} )\, \chi_{\sigma}^{\; a}
= ( \Gmt_{\mu\nu}^{ \sigma} - \Gmt_{\nu\mu}^{ \sigma} ) \, \chi_{\sigma}^{\; a} \equiv \Gmt_{[\mu\nu]}^{ \sigma}  \, \chi_{\sigma}^{\; a}  \, .
\ee

 \subsection{Quantum Dynamics of Space-time with Hidden Spinnic Gauge Symmetry }
 
Transmuting the spinnic gauge symmetry in Gravifield space-time into a hidden gauge symmetry and keeping the global Lorentz symmetry in the flat coordinate Minkowski space-time, we are able to reformulate the action for the theory of quantum gravity in terms of the space-time gauge field into the following general form 
\be
\label{action5}
S_{\chi}  & = & \int d^{4}x\, \chi\; \{\, \frac{1}{2}  [\,  \hat{\chi}^{\mu\nu} (\bar{\Psi} \chi_{\mu} i {\mathsf D }_{\nu}   \Psi  + \bar{\psi} \chi_{\mu}  i \td_{\nu}   \psi )  + H.c. \, ]  - y_s M_S \bar{\psi} \psi +  {\cal L}'(x)  \nonumber \\
& - & \frac{1}{4}  \hat{\chi}^{\mu\mu'} \hat{\chi}^{\nu\nu'} [\, {\cal F}^I_{\mu\nu} {\cal F}^{I}_{\mu'\nu'}  + W_{\mu\nu} W_{\mu'\nu'} + g_s^{-2} {\mathsf R}_{\mu\nu \rho}^{\sigma}  {\mathsf R}_{\mu'\nu' \rho'}^{\sigma'} \frac{1}{2}( \chi_{\sigma\sigma'}\chih^{\rho\rho'} - \eta^{\rho}_{\sigma'}\eta^{\rho'}_{\sigma} )   \, ]    \nonumber \\
& + &   \alpha_W M_S^2 \hat{\chi}^{\mu\mu'}  [  \frac{1}{4}  \hat{\chi}^{\nu\nu'} \chi_{\sigma\sigma'}  \Gmt_{[\mu\nu]}^{ \sigma}  \Gmt_{[\mu'\nu']}^{ \sigma'} + \frac{1}{2}( 3+ \frac{1}{\alpha_W}) g_w^2 w_{\mu}w_{\mu'} \, ] \nonumber \\
& - & \frac{1}{4}  \hat{\chi}^{\mu\mu'} \hat{\chi}^{\nu\nu'}(\, \chi_{\sigma\sigma'} \hat{\chi}^{\rho\rho'}   -  \eta_{\sigma}^{\rho} \eta_{\sigma'}^{\rho'} \, )  \ob_{\rho}\ob_{\rho'} \Gmt_{[\mu\nu]}^{ \sigma}  \Gmt_{[\mu'\nu']}^{ \sigma'} \nonumber \\ 
& + &   \alpha_E M_S^2 (\, 1 - \hat{\chi}^{\rho\sigma} \ob_{\rho}\ob_{\sigma}/\alpha_E M_S^2 \, )   {\mathsf R}   -  \lambda_s M_S^4    \nonumber \\
& + & g_s^{-1} (\td_{\sigma'} \ob_{\sigma}   - g_s \ob_{\sigma'} \ob_{\sigma}  \, )  \chih^{\mu\sigma'} 
(\, \chih^{\nu\rho} {\mathsf R}_{\mu\nu\rho}^{\sigma} + \chih^{\nu\sigma} {\mathsf R}_{\mu\nu} \, )    \nonumber \\
& + & g_s^{-1} \frac{1}{2} ( \chi_{\sigma\sigma'}\chih^{\rho\rho'} - \eta^{\rho}_{\sigma}\eta^{\rho'}_{\sigma'} )   \hat{\chi}^{\mu\mu'} \hat{\chi}^{\nu\nu'}  {\mathsf R}_{\mu\nu\rho}^{\sigma} \Gmt_{[\mu'\nu']}^{ \sigma'}  \ob_{\rho'} -  \hat{\chi}^{\mu\mu'} \hat{\chi}^{\nu\nu'}  \Gmt_{[\mu\nu]}^{ \sigma}  \ob_{\mu'} \td_{\nu'}\ob_{\sigma}  \,    \nonumber \\
& + & 6 \alpha_E g_s M_S^2  (\,  \hat{\chi}^{\mu\nu}  \td_{\mu}\ob_{\nu}  + g_s \hat{\chi}^{\mu\nu} \ob_{\mu}\ob_{\nu}  \, )  -3 g_s^2  (\, \hat{\chi}^{\mu\nu} \ob_{\mu}\ob_{\nu} )^2   \nonumber \\
& - & 2 \hat{\chi}^{\mu\mu'} \hat{\chi}^{\nu\nu'}  (\td_{\mu} \ob_{\nu})( \td_{\mu'}  \ob_{\nu'}) - 
(\chih^{\mu\nu} \td_{\mu} \ob_{\nu} )^2  \nonumber \\
 & + & 2 g_s \hat{\chi}^{\mu\nu}  \partial_{\mu}(\hat{\chi}^{\rho\sigma}\ob_{\rho} \ob_{\sigma}) \ob_{\nu} 
 - 4g_s \hat{\chi}^{\mu\nu} (\td_{\mu}\ob_{\nu} )\, \hat{\chi}^{\rho\sigma}\ob_{\rho} \ob_{\sigma}   \, \}
\ee
with the definitions
\be
& & \td_{\mu} \ob_{\nu} \equiv \partial_{\mu}\ob_{\nu} - \Gm_{\mu\nu}^{\sigma} \ob_{\sigma}\, ,  \qquad {\mathsf R} = \hat{\chi}^{\mu\nu}  {\mathsf R}_{\mu\nu}  = R + \Rmt  \, , \nonumber  \\
& & \Gm_{\mu\nu}^{\sigma}(x) \equiv  \Gamma_{\mu\nu}^{\sigma}(x)  +  \Gmt_{\mu\nu}^{\sigma}(x) \, , \qquad \Rm_{\mu\nu\rho}^{\sigma} \equiv  R_{\mu\nu\rho}^{\sigma} +  \Rmt_{\mu\nu\rho}^{\sigma}   \, .
\ee
Where we have defined the rank-2 tensor field strength $\Rm_{\mu\nu}$ as
\be
& &  \Rm_{\mu\nu} \equiv - \Rm_{\rho\mu\nu}^{\;\sigma} \eta_{\sigma}^{\rho}  = \Rm_{\mu\sigma\nu}^{\;\sigma}  =  
R_{\mu\nu}  +   \Rmt_{\mu\nu} \, , \nonumber \\
& & R_{\mu\nu}  = \partial_{\mu} \Gamma_{\sigma\nu}^{\sigma} - \partial_{\sigma} \Gamma_{\mu\nu}^{\sigma} -
\Gamma_{\mu\nu}^{\lambda} \Gamma_{\sigma\lambda}^{\sigma} + \Gamma_{\mu\lambda}^{\sigma}\Gamma_{\nu\sigma}^{\lambda} = R_{\nu\mu}\, , \quad \Gamma_{\sigma\nu}^{\sigma}  = \partial_{\nu} \ln \chi \, , \nonumber \\
 & & \Rmt_{\mu\nu}  =  \nabla_{\mu} \Gmt_{\sigma\nu}^{\sigma} - \nabla_{\sigma} \Gmt_{\mu\nu}^{\sigma}  -
\Gmt_{\mu\nu}^{\lambda} \Gmt_{\sigma\lambda}^{\sigma} + \Gmt_{\sigma\nu}^{\lambda} \Gmt_{\mu\lambda}^{\sigma} \, , \; \quad \Gmt_{\sigma\nu}^{\sigma}  \equiv \Gmt_{\nu} = \frac{1}{2} \chi_{\nu a} \chih_b^{\;\sigma} g_s \ot_{\sigma }^{ab} \, .
\ee
In the above action with hidden spinnic and scaling gauge symmetry and global Lorentz and scaling invariance, the singlet fermion $\psi(x)$, the scaling gauge field $w_{\mu}(x)$ and the space-time gauge field $\Gmt_{[\mu\nu]}^{\sigma}$ become massive in the Einstein-type basis after gauge symmetry breaking. The bosonic gravitational  interactions are described by the Goldstone-like Gravimetric field $\chi_{\mu\nu}(x)$ and the antisymmetric space-time gauge field $\Gmt_{\mu\nu}^{\sigma}$.  

It is seen that only the term given by the scalar tensor field strength $R$ characterizes the Einstein theory of general relativity. Obviously,  Einstein theory of general relativity  is expected as an effective low energy theory by integrating the high energy contributions into the renormalized coupling constants and quantum fields.

\subsection{ Gravity Equation Beyond Einstein's Equation of General Relativity} 

In terms of the hidden gauge formalism, the gravity equation Eq.(\ref{GE2}) can be written as
\be \label{GE21}
 \td_{\rho} {\cal G}^{\;\rho}_{ \mu\nu}  &  = &   {\cal T}_{\mu\nu} \,  ,  \qquad 
 {\cal G}^{\;\rho}_{ \mu\nu} = {\cal G}^{\;\rho\sigma}_{ \mu}\chi_{\sigma\nu}\, ,\;\, {\cal T}_{\mu\nu} =  {\cal T}_{\mu}^{\;\, \sigma} \chi_{\sigma\nu}
 \ee
 with the definition for the gauge invariant covariant derivative
 \be
  & & \td_{\rho} {\cal G}_{\mu\nu}^{\; \rho}  = ( \partial_{\rho} - g_w W_{\rho} ) {\cal G}_{\mu\nu}^{\; \rho} 
- \Gamma_{\rho\mu}^{\sigma} {\cal G}_{\sigma\nu}^{\; \rho} - \Gamma_{\rho\nu}^{\sigma} {\cal G}_{\mu\sigma}^{\; \rho} 
 - \Gmct_{\rho\mu}^{\sigma} {\cal G}_{\sigma\nu}^{\; \rho}  \, , \nonumber \\
 & & \Gmc_{\mu\nu}^{\sigma}(x)  =  \hat{\chi}_a^{\; \sigma} \td_{\mu} \chi_{\nu}^{\; a} = \hat{\chi}_a^{\; \sigma} (\partial_{\mu} \chi_{\nu}^{\; a} + g_s \Om_{\mu b}^{a} \chi_{\nu}^{\; b}  ) =  \Gamma_{\mu\nu}^{\sigma} (x)  + \Gmct_{\mu\nu}^{\sigma}(x)  \, , \nonumber \\
& & \Gamma_{\mu\nu}^{\sigma} (x)  \equiv \hat{\chi}_a^{\; \sigma} (\partial_{\mu} \chi_{\nu}^{\; a} + g_s \omega_{\mu b}^{a} \chi_{\nu}^{\; b}  )\, , \quad \Gmct_{\mu\nu}^{\sigma}(x) \equiv g_s \hat{\chi}_a^{\; \sigma} \Omt_{\mu b}^{a} \chi_{\nu}^{\; b} \, .
 \ee
where the spinnic gauge field $\Om_{\mu}^{ab}$ is the whole gauge field including the background field and decomposed into two parts $\Om_{\mu }^{ab} = \omega_{\mu }^{ab} + \Omt_{\mu }^{ab} $.
 
By comparing the symmetric part and antisymmetric part of  the gravity equation in Eq.(\ref{GE21}) 
\[ {\cal T}^{{\cal G}}_{\mu\nu} \equiv \td_{\rho} {\cal G}_{\mu\nu}^{\; \rho}  =  T^{{\cal G}}_{\mu\nu} + {\cal T}^{{\cal G}}_{[\mu\nu]} \, ,
 \quad  {\cal T}_{\mu\nu}\equiv G_{\mu\nu} + T_{\mu\nu} + {\cal T}_{[\mu\nu]} \, , \] 
we arrive at two types of gravity equation,
\be
& & G_{\mu\nu} =   - T_{\mu\nu} + T^{{\cal G}}_{\mu\nu} \, , \\
& & {\cal T}^{{\cal G}}_{[\mu\nu]}  =   {\cal T}_{[\mu\nu]} \, ,
\ee
where the first gravity equation is the extension to the Einstein's equation of general relativity, 
\be
& & 2\alpha_E M_S^2 \chi \, [\,  R_{\mu\nu} -   \frac{1}{2} \chi_{\mu\nu} \, R \, ]  \nonumber \\
& & \qquad  =  -[ \chi_{\mu\nu} T^{S} + T_{\mu\nu}^{F} - \chi_{\mu\nu} T^F +  T_{\mu\nu}^{R} - \frac{1}{2}\chi_{\mu\nu} T^R
+ T_{\mu\nu}^{G} - \frac{1}{4} \chi_{\mu\nu} T^G ] +  T_{\mu\nu}^W - \chi_{\mu\nu} T^W \nonumber \\
& & \qquad  + \alpha_W M_S^2 \chi \frac{1}{2} [\,   \hat{\chi}^{\rho\sigma}
 (\Gmct_{\rho\mu}^{\; \alpha} \Gmct_{[\sigma\nu\alpha]} + \Gmct_{\rho\nu}^{\; \alpha} \Gmct_{[\sigma\mu\alpha]}) 
 -g_w ( \Gmct_{\sigma\mu}^{\; \sigma} W_{\nu} + \Gmct_{\sigma\nu}^{\; \sigma} W_{\mu})  \, ] \, ,
\ee
and the second equation is a new type of gravitational equation 
\be
& & 2 \chi  \alpha_E M_S^2 \, ( \Rct_{\mu\nu} - \Rct_{\nu\mu} ) + \chi \alpha_W M_S^2   [\,  \hat{\chi}^{\rho\sigma}  ( \td_{\rho} \Gmct_{[\mu\sigma\nu]} - \td_{\rho} \Gmct_{[\nu\sigma\mu]}  )   + g_w ( \Gmct_{\sigma\mu}^{\; \sigma} W_{\nu} - \Gmct_{\sigma\nu}^{\; \sigma} W_{\mu}) \, ]   \nonumber \\
& &  = - \frac{1}{2}\chi  [\, 
i\bar{\Psi} ( \gamma_{\mu}{\mathcal D}_{\nu} - \gamma_{\nu}{\mathcal D}_{\mu} ) \Psi 
+ i\bar{\psi} (\gamma_{\mu} \td_{\nu} - \gamma_{\nu} \td_{\mu} ) \psi  + H.c.\,  ] - \chi \alpha_W M_S^2  g_w W_{\mu\nu} \, ,
\ee
which characterizes the twisting dynamics of spinnon. In above equations, we have used the following definitions
\be
T^S & = & \chi y_s M_S \bar{\psi} \psi  + \chi \lambda_s M_S^4  \, , \nonumber \\
T_{\mu\nu}^{F} & = & \frac{1}{4}\chi  [\, i\bar{\Psi} ( \gamma_{\mu}{\mathcal D}_{\nu} + \gamma_{\nu}{\mathcal D}_{\mu} ) \Psi 
+ i\bar{\psi} (\gamma_{\mu} \td_{\nu} + \gamma_{\nu} \td_{\mu} ) \psi  + H.c.\,  ] \, , \nonumber \\
T_{\mu\nu}^{R} & = &  \chi  \alpha_E M_S^2 \, ( \Rct_{\mu\nu} + \Rct_{\nu\mu} )  + \chi M_S^2 g_w^2 W_{\mu}W_{\nu}  \, , \nonumber \\
T_{\mu\nu}^{G} & = &  -  \chi \hat{\chi}^{\rho\sigma}  [\,  {\cal F}^I_{\mu\rho} {\cal F}^{I}_{\nu \sigma} 
+ {\cal W}_{\mu\rho} {\cal W}_{\nu \sigma} + g_s^{-2} {\cal R}_{\mu\rho \beta}^{\alpha}  {\cal R}_{\nu\sigma \beta'}^{\alpha'} \frac{1}{2}( \chi_{\alpha\alpha'}\chih^{\beta\beta'} - \eta_{\alpha\beta'}\eta_{\alpha'\beta} ) \, ]\, \nonumber \\
& + & \chi \alpha_W M_S^2    [\,  \hat{\chi}^{\rho\sigma}  \Gmct_{[\mu\rho]}^{ \alpha}  \Gmct_{[\nu\sigma\alpha]}  + 2 g_w^2 W_{\mu}W_{\nu}   \, ]  \, , \nonumber \\
T^F & = & T_{\mu\nu}^{F} \hat{\chi}^{\mu\nu} \, , \quad T^R  =  T_{\mu\nu}^{R} \hat{\chi}^{\mu\nu} \, , \quad  T^G  =  T_{\mu\nu}^{G} \hat{\chi}^{\mu\nu} \, , \quad \Gmct_{[\nu\sigma\alpha]}  \equiv \Gmct_{[\nu\sigma]}^{\beta} \chi_{\beta\alpha}  \, ,
\ee
and 
\be
& & T_{\mu\nu}^W  =  \alpha_W M_S^2 g_w \chi  \frac{1}{2} ( \td_{\mu} W_{\nu} + \td_{\nu} W_{\mu} ) \, , \quad T^W = \hat{\chi}^{\mu\nu} T_{\mu\nu}^W \, ,\nonumber \\
& & \td_{\mu} W_{\nu} =  ( \partial_{\mu} - g_w W_{\mu} ) W_{\nu} - \Gamma_{\mu\nu}^{\rho} W_{\rho} \, , \quad W_{\mu\nu} = \partial_{\mu}W_{\nu} - \partial_{\nu}W_{\mu}\, , \nonumber \\
& & \td_{\rho} \Gmct_{[\nu\sigma\mu]}  =  (\partial_{\rho} - g_wW_{\rho}) \Gmct_{[\nu\sigma\mu]}  
- \Gamma_{\rho\nu}^{\alpha} \Gmct_{[\alpha\sigma\mu]} - \Gamma_{\rho\sigma}^{\alpha} \Gmct_{[\nu\alpha\mu]}
- \Gamma_{\rho\mu}^{\alpha} \Gmct_{[\nu\sigma\alpha]}   - \Gmct_{\rho\mu}^{\; \alpha} \Gmct_{[\nu\sigma\alpha]} 
\ee
In comparison with the Einstein theory of general relativity, the combination of coupling parameter and scaling mass scale  is fixed to be
\be
2\alpha_E M_S^2 = \frac{1}{8\pi G} = \frac{1}{8\pi } M_P^2 
\ee
with $M_P$ the Planck mass.

\section{Concluding Remarks}

Some underlying principles alternative to the Einstein's general coordinate invariance have been proposed to establish the theory of quantum gravity. The main principle beyond Einstein's principle is that: theory of quantum gravity must be expressed to be coordinate independence and gauge invariance in Gravifield space-time. A bi-frame space-time has been initiated to describe the laws of nature. One frame space-time is the coordinate space-time depicted by the globally flat Minkowski space-time that acts as an inertial reference frame for the motions of fields, the other is the non-coordinate space-time characterized by the locally flat Gravifield space-time that functions as an interaction representation frame for the degrees of freedom of fields. This enables us to present a QFT description for the gravity by treating the gravitational force on the same footing as the electroweak and strong forces. 
 
When transmuting the Gravifield basis in Gravifield space-time into the coordinate basis in Minkowski space-time, we have obtained equations of motion for all quantum fields and derived basic conservation laws for all symmetries. The gravity equation alternative to Einstein equation of gravity has heen shown to be governed by the total energy-momentum tensor defined in the flat Minkowski space-time. The spinnic and scaling gauge symmetry breaking has been proposed to obtain a Lorentz invariant background Gravifield space-time. The exact solution for the background fields has been yielded to form a conformally flat Minkowski space-time as the background Gravifield space-time. The background Gravifield space-time has been shown to be characterized by a cosmic vector with a nonzero cosmological mass scale. In terms of the conformal proper time, such a background Gravifield space-time has a cosmological horizon that can only be reached in an infinitely large cosmic proper time and leads to a solution of inflationary universe. In the quantized gravitational interactions, we have demonstrate that massless graviton and massive spinnon are the physical degrees of freedom when the spinnic and scaling gauge symmetries are broken down to a background structure. It has been seen that there exist rich gravitational interactions with the background fields. The quantum effect has been shown to play a significant role on the inflation of the universe.  In fact, it is the quantum loop quadratic contributions that cause the breaking of global scaling symmetry and generate the inflation of early universe. The inflationary universe gets end when the evolving vacuum expectation value of the background scalar potential approaches to a global minimal. 

The Gravifield behaves as a Goldstone-like field that transmutes the local spinnic gauge symmetry into the global Lorentz symmetry, which makes the spinnic gauge field becomes a hidden gauge field. The bosonic gravitational interactions have been shown to be characterized by the Goldstone-like Gravimetric field and space-time gauge field, while the fermionic gravitational interactions must remain to be described by the Gravifield. The Einstein theory of general relativity has been shown to be an effective low energy theory, its effect on the gauge couplings have been studied in Ref.\cite{TW}, which leads to a power law running.  We have demonstrated that there are in general two types of gravity equation, one is the extension to the Einstein's equation of general relativity that describes the dynamics of graviton and space-time, and the other is a new type of gravitational equation that characterizes the dynamics of spinnon and twisting effect.

\centerline{{\bf Acknowledgement}}

I would like to dedicate this talk to Professor Huan-Wu Peng for the 100 anniversary of the birth. The author is grateful to many colleagues for valuable conversations and discussions during the International Conference on Gravitation and Cosmology/the fourth Galileo-Xu Guangqi Meeting, held at KITPC/ITP-CAS and UCAS on May 4-8, 2015, for celebrating the 100th anniversary of Albert Einstein's presentation of the Theory of General Relativity. This work was supported in part by the National Science Foundation of China (NSFC) under Grant \#No. 11475237, No.~11121064, No.~10821504 and also by the CAS Center for Excellence in Particle Physics (CCEPP).
   

\end{document}